\documentclass[12pt,a4paper]{report}
\usepackage[italian, english]{babel}
\usepackage{newlfont}
\usepackage{color}
\usepackage{setspace}

\usepackage[toc, page]{appendix}
\usepackage{bm}
\usepackage{amsfonts}
\usepackage{graphicx}
\usepackage{amsmath}
\usepackage{amssymb}
\usepackage{amsthm}
\usepackage[justification=centerlast, font=small, labelfont=bf]{caption}
\graphicspath{{./dati//}}
\usepackage{float}
\usepackage[colorlinks = true, linkcolor = blue, urlcolor  = blue, citecolor = blue, anchorcolor = blue]{hyperref}

\hypersetup{
	pdfpagemode={UseOutlines},
	bookmarksopen,
	pdfstartview={FitH},
	colorlinks,
	linkcolor={black},
	citecolor={black},
	urlcolor={black}}
\usepackage{caption}
\usepackage{subcaption}
\usepackage{siunitx}

\usepackage{booktabs}
\usepackage[bottom]{footmisc}
\usepackage{gensymb}
\usepackage{wasysym}
\usepackage{dcolumn}

\usepackage[english]{babel}
\usepackage[utf8]{inputenc}
\usepackage{multirow}
\usepackage{fixltx2e}
\begin{document}
\begin{titlepage}
%
%
% UNA VOLTA FATTE LE DOVUTE MODIFICHE SOSTITUIRE "RED" CON "BLACK" NEI COMANDI \textcolor
%
%
\begin{center}
{{\Large{\textsc{Alma Mater Studiorum $\cdot$ Universit\`a di Bologna}}}} 
\rule[0.1cm]{15.8cm}{0.1mm}
\rule[0.5cm]{15.8cm}{0.6mm}
\\\vspace{3mm}
%
% PER LA LAUREA TRIENNALE TOGLIERE "MAGISTRALE"
%
{\small{\bf Scuola di Scienze \\
Dipartimento di Fisica e Astronomia\\
Corso di Laurea Magistrale in Fisica}}

\end{center}

\vspace{40mm}

\begin{center}
{\LARGE
\textbf{sEMG-based}
\textbf{Hand Gesture Recognition}

\vspace{4mm}

\textbf{with Deep Learning}
} %\\
\end{center}

\vspace{30mm} \par \noindent

%\ begin{center}\textcolor{black}{
%
% INSERIRE IL TITOLO DELLA TESI
%
%{\LARGE{\textbf{sEMG-based\\Hand Gesture Recognition\\with Deep Learning}}}\\
%}\end{center}

%\vspace{30mm} \par \noindent

\begin{minipage}[t]{0.47\textwidth}
%
% INSERIRE IL NOME DEL RELATORE CON IL RELATIVO TITOLO DI DOTTORE O PROFESSORE
%
{\large{\bf Relatore: \vspace{2mm}\\\textcolor{black}{
Prof. Daniel Remondini}\\\\
%
% INSERIRE IL NOME DEL CORRELATORE CON IL RELATIVO TITOLO DI DOTTORE O PROFESSORE
%
% SE NON AVETE UN CORRELATORE CANCELLATE LE PROSSIME 3 RIGHE
%
\textcolor{black}{
\bf Correlatori:
\vspace{2mm}\\
Dott. Simone Benatti\\
Dott. Francesco Conti\\
Dott. Manuele Rusci\\\\}}}
\end{minipage}
%
%\hfill
%
\begin{minipage}[t]{0.47\textwidth}\raggedleft \textcolor{black}{
{\large{\bf Presentata da:
\vspace{2mm}\\
%
% INSERIRE IL NOME DEL CANDIDATO
%
Marcello Zanghieri}}}
\end{minipage}

%\vspace{40mm}
%\vspace{20mm}

\begin{center}
%
% INSERIRE L'ANNO ACCADEMICO
%
Anno Accademico \textcolor{black}{2017/2018}
\end{center}

\end{titlepage}

%\title{
%\LARGE{Cdl fisica, curriculum applicata}\\
%\LARGE{Tesi magistrale}\\
%\vspace{2.5cm}
%\textbf{sEMG-based hand gesture recognition with deep learning: addressing postural and temporal variability by means %of different training strategies}
%}
%\date{Marzo 2019}
% %\author{Gruppo 1:\\
% %Bravaglieri Lisa, Giovannelli Ilaria, \\Madsen Mads Holst Aagaard, Magkos Sotirios, \\Nurrito Eugenio, Tavernari
% %Daniele.\\\\
% %\large{Università di Bologna, Dipartimento di Fisica e Astronomia}}
%\author{Marcello Zanghieri\\\\\\\\\\
% \large{Università di Bologna, Dipartimento di Fisica e Astronomia}}
%\maketitle

% % % % % % % % % % % % % % % % % % % % % % % % % % % % % % % % % % % % % % % % % % % % % % % % % % % % % % % % % % % % % % % % % % % % % % % % % % % % % % % % % % % % % % % % % % % % % % % % % % % % % % % % % % % % % % % % % % % % % % % % % % % % % % % % % % % % % % % % % % % % % % % % % % % % % % % % % % % % % % % % % % % % % % % % % % % % % % % % %

\begin{abstract} % Abstract

Hand gesture recognition based on surface electromyographic (sEMG) signals is a promising approach for the development of Human-Machine Interfaces (HMIs) with a natural control, such as intuitive robot interfaces or poly-articulated prostheses. However, real-world applications are limited by reliability problems due to motion artifacts, postural and temporal variability, and sensor re-positioning.

This master thesis is the first application of deep learning on the Unibo-INAIL dataset, the first public sEMG dataset exploring the variability between subjects, sessions and arm postures, by collecting data over 8 sessions of each of 7 able-bodied subjects executing 6 hand gestures in 4 arm postures. In the most recent studies, the variability is addressed with training strategies based on training set composition, which improve inter-posture and inter-day generalization of classical (i.e.\ non-deep) machine learning classifiers, among which the RBF-kernel SVM yields the highest accuracy.

The deep architecture realized in this work is a 1d-CNN implemented in Pytorch, inspired by a 2d-CNN reported to perform well on other public benchmark databases. On this 1d-CNN, various training strategies based on training set composition were implemented and tested.

Multi-session training proves to yield higher inter-session validation accuracies than single-session training. Two-posture training proves to be the best postural training (proving the benefit of training on more than one posture), and yields 81.2\% inter-posture test accuracy. Five-day training proves to be the best multi-day training, and yields 75.9\% inter-day test accuracy. All results are close to the baseline. Moreover, the results of multi-day trainings highlight the phenomenon of user adaptation, indicating that training should also prioritize recent data.

Though not better than the baseline, the achieved classification accuracies rightfully place the 1d-CNN among the candidates for further research.

\end{abstract}

\begingroup
\selectlanguage{italian}
\begin{abstract}% Sommario %\vspace{-3mm}

Il riconoscimento di movimenti della mano basato su segnali elettromiografici di superficie (sEMG) è un approccio promettente per lo sviluppo di interfacce uomo-macchina (HMI) a controllo naturale, quali interfacce robot intuitive o protesi poliarticolate. Tuttavia, le applicazioni \textit{real-world} sono limitate da problemi di affidabilità dovuti ad artefatti di movimento, variabilità posturale e temporale, e riposizionamento dei sensori.

Questa tesi magistrale è la prima applicazione dell'apprendimento profondo (\textit{deep learning}) al dataset Unibo-INAIL, il primo dataset sEMG pubblico che esplora la variabilità tra soggetti, sessioni e posture del braccio, raccogliendo dati da 8 sessioni per ciascuno di 7 soggetti integri che eseguono 6 movimenti della mano in 4 posture del braccio. Negli studi più recenti, la variabilità è affrontata con strategie di addestramento basate sulla composizione del \textit{training set}, che migliorano la generalizzazione inter-postura e inter-giorno di classificatori basati sull'apprendimento automatico (\textit{machine learning}) classico (ossia non profondo), fra i quali la RBF-kernel SVM dà l'accuratezza più alta.

L'architettura profonda realizzata in questo lavoro è una 1d-CNN implementata in PyTorch, ispirata a una 2d-CNN con buone prestazioni dimostrate su altri dataset pubblici di riferimento. Su questa 1d-CNN, si implementano e testano varie strategie di addestramento basate sulla composizione del \textit{training set}.

%La strategia di addestramento mono-sessione raggiunge un'accuratezza di validazione intra-sessione del 94.5\%, ma si degrada a 80.6\% in validazione inter-postura e a 66.9\% (per il giorno 1) in validazione inter-giorno. La variabilità inter-giorno è quindi maggiore della variabilità inter-postura.
%A possible reason is that, on each day, the data of the 4 arm postures were collected without repositioning the sensors.

L'addestramento multi-sessione si dimostra produrre accuratezze di validazione inter-sessione più alte rispetto all'addestramento mono-sessione. L'addestramento bi-postura si dimostra la miglior strategia posturale (dimostrando il beneficio di addestrare su più di una postura), e produce un'accuratezza di test inter-postura dell'81.2\%. L'addestramento su cinque giorni si dimostra la miglior strategia multi-giorno, e produce un'accuratezza di test inter-giorno del 75.9\%. Tutti i risultati sono vicini alla \textit{baseline}. Inoltre, i risultati degli addestramenti multi-giorno evidenziano il fenomeno dell'adattamento dell'utente, indicando che l'addestramento deve privilegiare dati recenti.

Benché non migliori della \textit{baseline}, le accuratezze ottenute pongono a buon diritto la 1d-CNN tra i modelli candidati per la ricerca futura.

\end{abstract}
\endgroup

% % % % % % % % % % % % % % % % % % % % % % % % % % % % % % % % % % % % % % % % % % % % % % % % % % % % % % % % % % % % % % % % % % % % % % % % % % % % % % % % % % % % % % % % % % % % % % % % % % % % % % % % % % % % % % % % % % % % % % % % % % % % % % % % % % % % % % % % % % % % % % % % % % % % % % % % % % % % % % % % % % % % % % % % % % % % % % % % %

\pagenumbering{roman}
\tableofcontents
%\pagenumbering{arabic}

% % % % % % % % % % % % % % % % % % % % % % % % % % % % % % % % % % % % % % % % % % % % % % % % % % % % % % % % % % % % % % % % % % % % % % % % % % % % % % % % % % % % % % % % % % % % % % % % % % % % % % % % % % % % % % % % % % % % % % % % % % % % % % % % % % % % % % % % % % % % % % % % % % % % % % % % % % % % % % % % % % % % % % % % % % % % % % % % %

\chapter{Introduction}\label{introduction}
\pagenumbering{arabic}

Hand gesture recognition based on electromyographic (EMG) signals is an innovative approach for the development of human-computer interaction, the vast field whose aim is to implement human-computer interfaces (HMIs) and intuitive interaction devices. The research in this field is motivated by the need for intelligent devices able to extract information from data coming from sensors, operating in real time and under significant power, size and cost constraints. The wide range of applications of HMIs with EMG-based intuitive control includes (but is not limited to) robot interaction and industrial robot control, game or mobile interfaces, interactions for virtual environments, sign language recognition, rehabilitation, and control of poly-articulated prostheses \cite{Cheok2017, Farina2014, Meattin2018, Saponas2010, Hakonen2015}.

The electromyographic (EMG) signal is the biopotential generated by the ionic flow through the membrane of the muscular fibers during contraction, and is a major index of the muscular activity. EMG data can be acquired either with invasive or non-invasive measuring instruments. Invasive methods employ wire or needle electrodes, which penetrate the skin to reach the muscle of interest. On the contrary, surface electromyography (sEMG) is a non-invasive technique that uses surface electrodes applied on the surface of the skin \cite{Tsinganos2018}. In the HMI field, building gesture recognition on the analysis of sEMG signals is one of the most promising approaches, since non-invasiveness is an essential requirement for many types of HMIs.
% In the sEMG setup, APs can be detected by means of an instrumentation amplifier with the positive and negative terminals each one connected to two metal plates positioned on the skin surface: the sEMG signal is finally formed by the superposition of all the detected APs underlying the amplifier \cite{Milosevic2017}.

An open challenge in HMI design is the development of solutions based on a robust recognition approach. On the one hand, the implementation of devices showing high recognition capabilities in controlled environments raised industrial interest and led to the availability of commercial solutions based on the EMG-based interaction paradigm. On the other hand, in many real-world scenarios the adoption of EMG-based HMIs is still limited by reliability problems such as motion artifacts, postural and temporal variability, and issues caused by sensors re-positioning at each use.

Nowadays, research efforts focus on solving issues related to postural variability effects and long-term reliability. Such research efforts are being boosted by two factors. (1) The release of public EMG databases suitable for the analysis of variability, whose publication eases the creation of benchmarks for the research community; the Unibo-INAIL dataset studied in this master thesis was realized for this very purpose. (2) The increasing recourse to the deep learning, whose deep hierarchical approach, together with the entirely data-driven feature extraction, promises to speed up the search for effective representations able to empower the recognition models. % In particular, deep learning has gathered attention because its deep hierarchical approach, together with the entirely data-driven selection of discriminative features, have shown much better results in terms of generalization in other different but analogous fields, raising the interest also for sEMG-based geture recognition.

\section{EMG-based HMIs and generalization issues}
The Electromyogram (EMG) is the biopotential signal resulting from muscular activity, and is a major index thereof. It can be sensed by means of non-invasive surface electrodes, giving rise to the surface EMG (sEMG) signal. The processing of sEMG signals is a promising approach for the implementation of non-invasive EMG-based Human-Machine Interfaces.

However, the current state-of-the-art has to cope with challenging issues. The sEMG signal is severely affected by many factors, such as differences between subjects, fatigue, user adaptation and the variability introduced from the re-positioning of electrodes at each data collection session. These issues limit long-term use and reliability of the devices relying on EMG analysis.

% PARAGRAFO MODIFICATO, PERCHE' RIPETUTO NEL SOA
In the machine learning framework, these variability factors can be modeled with the concept of data sources, i.e.\ data subsets coming from different distributions (the concept of multi-source data is defined in Subsection \ref{multisource}). Identifying multiple sources makes machine learning on EMG data a challenging task, where the ambition is to implement classifiers capable of good inter-source generalization, e.g. in inter-posture, inter-session or inter-subject scenarios. Up to now, the classification accuracy with Leave-One-Subject-Out cross-validation (LOSOCV) is still much lower than that attained with Within-Subject cross-validation (WSCV) \cite{Du2017}.

\section{Deep Learning revolution}
At present, an increasingly important role in sEMG-based gesture recognition (and in human-computer interaction in general) is being played by deep learning. Existing deep learning architectures are mainly based on two kinds of architecture: the Convolutional Neural Network (CNN), able to capture spatial information of the signal, and the Recurrent Neural Network, which allows to exploit the sequential nature of the data. This master thesis is exclusively focused on the CNN architecture, because CNNs are (1) easier to train, and (2) more suitable for deployment in embedded hardware, hence making them more attractive for this work. On the other hand, it must be noted that RNNs are typically more accurate on time-domain data, so the choice between the two models is not clear cut.

In the framework of sEMG-based gesture recognition based on classical machine learning, the pipeline typically consists of data acquisition, data preprocessing, feature extraction, feature selection, model definition and inference \cite{Tsinganos2018, Hu2018}. The reason behind the new deep learning perspective for sEMG-based gesture recognition is that it mitigates the strong need for feature extraction, feature selection and parameter tuning, which strongly rely on specific domain knowledge. The advantage of a deep architecture is its ability to incorporate feature learning: a consistent part of the traditional pipeline can be entrusted to the training of the algorithm, which has enough capacity to learn effective feature representations on its own. This mitigates the reliance on rigid combinations of preprocessing steps and precise sets of hopefully discriminative features.

The deep learning approach has already started to speed up the research in this field. In particular, the CNN-based sEMG gesture recognition has been studied by Atzori et al.\ in \cite{Atzori2016}, achieving comparable performance with traditional methods on the Non-Invasive Advanced Prosthetics (NinaPro) database. Another CNN architecture with adaptive feature learning improving inter-subject generalization has been proposed by Park and Lee in \cite{ParkLee2016}. Geng et al.\ \cite{Geng2016, Du2017-1} presented a new CNN architecture for instantaneous sEMG images on three sEMG benchmark datasets. Du et al.\ \cite{Du2017-2} designed a semi-supervised deep CNN which also exploits the auxiliary information of a data glove: classification performance is improved by training the auxiliary task of regression of glove signal.

\section{Purpose of this master thesis}
The purpose of this master thesis is to apply deep learning for the first time to the Unibo-INAIL dataset, the first public dataset of surface electromyographic signals (sEMG) exploring the impact of combined postural and temporal variabilities on myoelectric hand gesture recognition \cite{Milosevic2017}. In particular, the deep learning architecture used is a one-dimensional Convolutional Neural Network (1d-CNN), which performs convolutions over the time dimension.

This is a novel approach since the application of deep learning yields new knowledge about the performance of machine learning on this dataset, which was previously analysed only with algorithms belonging to classical (i.e.\ non-deep) machine learning, applied only to instantaneous signal values. The CNN architecture used in this work is an adaptation of the 2d-CNN module of the hybrid CNN-RNN architecture proposed by Hu et al.\ in \cite{Hu2018}. The evaluation of performance is made in such a way to directly compare deep learning and classical machine learning on each of the intra- and inter-session scenarios available as baseline for the various training and validation sets compositions.

At the dataset's state-of-the-art, the variability is addressed with training strategies that improve inter-posture and inter-session generalization of classical (i.e.\ non-deep) machine learning classifiers, among which the RBF-kernel SVM yields the highest accuracy. The deep model implemented and tested is a CNN architecture reported to perform well on other public benchmark databases. On this CNN, the state-of-the-art training strategies are implemented and tested.

% NON mettere l'anticipazione dei risultati.
% Accuracies show an inter-posture deterioration up to 16\% and an inter-day deterioration up to 20\%. However, multi-posture and multi-day training strategies improve the performances. Two-posture training proves to be the best postural training strategy, and yields 81.2\% inter-posture test accuracy. Five-day training proves to be the best multi-day training strategy, and yields 75.9\% inter-day test accuracy. These accuracies are close to the baseline, thus equalling SoA. Moreover, the results of multi-day trainings highlight a progressive trend, interpreted as user adaptation, indicating that (re)training strategies should prioritize the availability of recent data.
% The interpretation of these results is that the generalization accuracy attained is the maximum attainable on the Unibo-INAIL dataset with the chosen preprocessing.

\section{Thesis structure}

The remainder of this master thesis is structured as follows:

\begin{itemize}\itemsep0em

\item[$\circ$] Chapter 2 introduces the fundamentals of surface electromyography (from muscular activation potential to signal collection at the electrodes), compares the classical machine learning framework and the deep learning framework of sEMG-based hand gesture recognition, and summarizes the state of the art regarding the classification robustness against the signal variability factors (both in general and on the Unibo-INAIL dataset);

\item[$\circ$] Chapter 3 introduces the fundamentals of deep learning, of convolutional neural networks and of the architectural features used in this master thesis;

\item[$\circ$] Chapter 4 illustrates the materials and methods of this work: the Unibo-INAIL dataset, the adopted preprocessing and pipeline, the CNN architecture used (together with all the training settings), and the training strategies implemented and tested;

\item[$\circ$] Chapter 5 describes the scripts developed and how they exploit the main packages of the open source deep learning platform PyTorch;

\item[$\circ$] Chapter 6 presents the results obtained for the various training strategies;

\item[$\circ$] Chapter 7 exposes the conclusions and outlines the future work.

\end{itemize}

% % % % % % % % % % % % % % % % % % % % % % % % % % % % % % % % % % % % % % % % % % % % % % % % % % % % % % % % % % % % % % % % % % % % % % % % % % % % % % % % % % % % % % % % % % % % % % % % % % % % % % % % % % % % % % % % % % % % % % % % % % % % % % % % % % % % % % % % % % % % % % % % % % % % % % % % % % % % % % % % % % % % % % % % % % % % % % % % %
\newpage
\chapter{Surface Electromyography and sEMG-based gesture recognition}\label{EMG}

Surface Electromyography is the field that studies the sensing, elaboration and uses of the surface-electromyographic (sEMG) signal, i.e.\ the electrical signal generated by contractions of skeletal muscles and sensed by non-invasive surface electrodes. For the development of Human-Machine Interfaces (HMIs), building gesture recognition on the analysis of sEMG signals is one of the most promising approaches, since for many applications non-invasiveness is an essential requirement.

This chapter is structured as follows:
\begin{itemize}\itemsep0em
\item[$\circ$] Section 2.1 introduces the fundamental concepts of surface electromyography, from the muscular activation potentials to the signal collection at the electrodes;
\item[$\circ$] Section 2.2 is devoted to sEMG-based gesture recognition: it provides an overview of the techniques and results of both the classical machine learning framework and the deep learning framework, and summarizes the state of the art regarding the search for classifiers that are robust against the signal variability factors, both in the general field and on the Unibo-INAIL dataset.
\end{itemize}

\section{Surface Electromyography}\label{sEMG}

Electromyography \cite{DeLuca2006, Tassinary} is the discipline that studies the detection, analysis, and applications of the electromyographic (EMG) signal, the electrical signal generated by muscles in correspondence with contractions. In particular, surface electromyography (sEMG) is a non-invasive technique for measuring and analysing the EMG signal of skeletal muscles by means of surface electrodes.

The electromyographic (EMG) signal is the bio-electric potential that arises from the current generated by the ionic flow through the membrane of the muscular fibers. It is therefore a major index of the muscular activity. During a muscular contraction, the depolarization of the tissue cell membrane, caused by the flow of Na$^+$ and K$^+$ ions, propagates along the muscle fibers. The origin of the potential is the electrical stimulus that starts from the central nervous system, and passes through the motor neurons (motoneurons) innervating the muscular tissue, giving rise to the Action Potentials (APs).

EMG data can be acquired either with invasive or non invasive measuring instruments. Invasive methods employ wire or needle electrodes, which penetrate the skin to reach the muscle of interest. On the contrary, surface electromyography (sEMG) is a non-invasive technique that uses surface electrodes that operate on the surface of the skin \cite{Tsinganos2018}. For many of human-machine interfaces, non-invasiveness is an indispensable requirement. In the sEMG setup, APs can be detected by means of an instrumentation amplifier with the positive and negative terminals connected to two metal plates positioned on the skin surface: the sEMG signal results from by the superposition of all the detected APs underlying the amplifier \cite{Milosevic2017}.

The EMG signal amplitude depends on the size and distance of the muscles underlying the electrodes, and typically ranges from $\SI{10}{\micro\volt}$ to $\SI{10}{\milli\volt}$. The EMG signal bandwidth stays within $\SI{2}{\kilo\hertz}$. The noise sources affecting the EMG signal are many, the most important being motion artifacts, floating ground noise and crosstalk. A major source of interference is Power Line Intereference (PLI) \cite{Tomasini2016}, which is due to the capacitive coupling between the body and the surrounding electrical devices and power grid, and is common to most biomedical signals. Despite having nominal main frequency \SI{50}{\Hz} in Europe and \SI{60}{\Hz} in USA, PLI is non-stationary and can present variations in frequency (up to $\pm\SI{2}{\Hz}$) and variations in amplitude (depending on instrumentation and environment), both mainly originating from the AC power system. Though beyond the scope of this master thesis, the development of filters and removal algorithms to reject PLI effectively is an active field of research.

\subsection{Muscular activation potentials}

% The action potential (AP) is the electrical signal that accompanies the mechanical contraction of a single cell when stimulated by an electrical current (neural or external) [lo, 17, 18, 19, 20, 211. It is caused by the flow of sodium (Nu+), potassium (K+), chloride (CZ-)a,n d other ions across the cell membrane. The action potential is the basic component of all bioelectrical signals. It provides information on the nature of physiological activity at the single-cell level.

Motor units are the basic functional units of a muscle, and muscle fibers are innervated motor units. When activated, motor units generate a Motor Unit Action Potential (MUAP). MUAPs can be generated in sequence by repeated activations, giving rise to MUAP Trains (MUAPTs). MUAPTs are possible as long as the muscle is able to generate force.

A convenient elementary model of MUAPs can be built with the tools of linear response theory \cite{Rangayyan2015}. In the linear response model, represented in Figure \label{img_EMG_linres}, a MUAPT can be fully described with two mathematical entities: the sequence of instants of the $n$ pulses of the train  $\{t_i\}_{i = 1, \cdots, n}$, and the MUAP linear response function $h(t)$, accounting for the shape of the MUAP and also called the MUAP waveform. Interpulse intervals are defined as $\{\Delta_i = t_{i + 1} - t_i\}_{i = 1, \cdots, n - 1}$, i.e.\ the time intervals between consecutive MUAPs. Each impulse is a Dirac delta impulse $\delta(t)$ (using dimensionless quantities), and the impulse train $\bar{\delta}(t)$ is given by their sum:
\begin{align}
\bar{\delta}(t) = \sum_{i = 1}^n \delta(t - t_i).
\end{align}
The expression $u(t)$ that describes the MUAPT as a function of time is obtained applying the kernel $h$ on the impulse train $\bar{\delta}(t)$:
\begin{align}
u = h \ast \bar{\delta}
\end{align}
which more explicitly is
\begin{align}
u(t) = (h \ast \bar{\delta})(t) &= \int_{-\infty}^{+\infty} \bar{\delta}(t - t') h(t') \textrm{d}t' =\\
%&= \int_{0}^{+\infty} \bar{\delta}(t - t') h_\textrm{delayed}(t') \textrm{d}t' =\\
%&= \int_{0}^{+\infty} \sum_{i = 1}^n \delta(t - t_i - t') h_\textrm{delayed}(t') \textrm{d}t' =\\
&=\sum_{i = 1}^n h(t - t_i)
\end{align}
where $\ast$ denotes convolution.

\begin{figure}[H]
  \centering
  \makebox[\textwidth][c]{\includegraphics[width=1\textwidth]{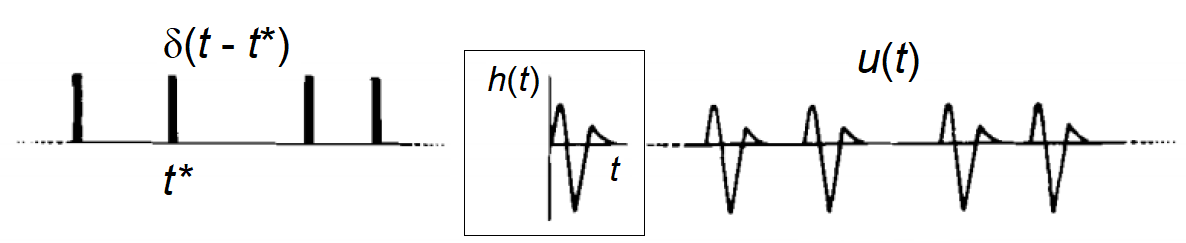}}
  \caption{Linear response model of the Motor Unit Action Potential Train (MUAPT). Image adapted from \cite{Rangayyan2015}.}
  \label{img_linres}
\end{figure}

\begin{figure}[H]
  \centering
  \makebox[\textwidth][c]{\includegraphics[width=0.9\textwidth]{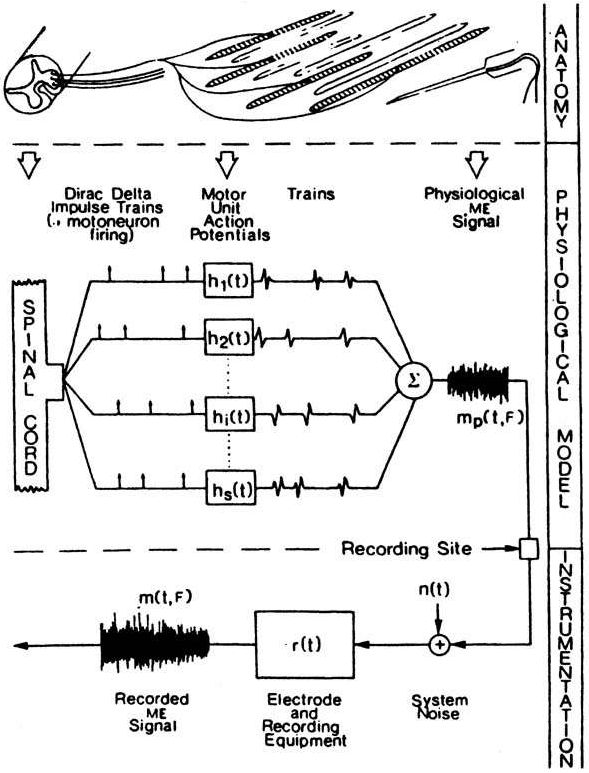}}
  \caption{Scheme of the potential from Activation Potential (AP) to the recorded EMG signal. Image from \cite{Rangayyan2015}.}
  \label{img_EMG_scheme}
\end{figure}

More precisely, MUAP values depend also on the force $F$ generated by the muscle: $u = u(t; F)$. The value of the EMG signal $m(t; F)$ sensed at time $t$ depends on the generated force $F$ as well, and is given by the sum of all the MUAPTs $u_\textrm{unit}(t)$ generated by the probed motor units:
\begin{align}
m(t; F) = \sum_{\textrm{unit} \in U} u_\textrm{unit}(t; F)
\end{align}
where $U$ is the set of the probed motor units. A comprehensive scheme following the potential from the AP to the recorded EMG signal is shown in Figure \ref{img_EMG_scheme}.

In both invasive and surface electromyography, the observed MUAP waveform depends on the relative position between the electrode and the active muscle fibers. This relative position may not be constant over time. The MUAP waveform and the EMG signal are also affected by any significant biochemical change in the muscle tissue: this is the case of muscle fatigue, which is a known source of variability for the EMG signal.

\section{sEMG-based gesture recognition}
sEMG-based gesture recognition is a promising technique for the development of Human-Machine Interfaces (HMIs). On the one hand, sEMG has the virtue of being non-invasive, which is essential for many HMIs. On the other hand, entrusting recognition, i.e.\ signal classification, to automated learning is a successful method for circumventing the complexity of the task by removing the need of a complete physiological understanding of the underlying motor functions.

Automated learning has produced advances with a rich variety of approaches. Whereas the central goal of gesture recognition is signal classification, many other machine learning techniques have been successfully exploited as auxiliary tasks to assist classification at either training or evaluation time, increasing classification accuracy. Examples of auxiliary tasks (detailed in the following subsections) are force estimation by means of regression, and semi-supervised and unsupervised methods for calibration.

However, classical machine learning still requires field-specific knowledge, such as tested and established feature extracting procedures to maximize discriminative power, recourse to the suitable signal domain (time domain, frequency domain, or time-frequency domain), and empirical expertise about the portability of preprocessing and extraction procedures across datasets acquired with different experimental setups.

This need for field-specific knowledge can be overcome by migrating to deep learning algorithms, capable of autonomous feature learning, i.e.\ to learn good feature representations autonomously, provided sufficient data. The success of deep-learning models has even driven the ambition of reversing the perspective: once trained a satisfactory deep classifier, model explainability can be leveraged to shed light a posteriori on the feature representations, or even on the physiological understanding, lacked a priori. Steps toward demonstrating the feasibility of this approach have already been moved in the related field of electroencephalogram (EEG)-based gesture recognition for Brain-Computer Interfaces (BCIs): for instance, Nurse et al.\ in \cite{Nurse2015, Nurse2016}, have shown correspondences between channel-time domain filters learned by the first layer of CNNs and known activation patterns of brain regions.

\subsection{Classical Machine Learning approach}
Classical Machine Learning (classical ML), also referred to as traditional or conventional machine learning, is the broad class of non-deep algorithms which includes $k$-Nearest Neighbors ($k$NN), Support Vector Machine (SVM), Linear Discriminant Analysis (LDA), and Random Forests (RF) (all with the related kernel methods), and Multi-Layer Perceptron (MLP) with one hidden layer. In the framework of sEMG-based gesture recognition based on classical ML, the pipeline typically consists of data acquisition, data preprocessing, feature extraction, feature selection, model definition, and inference \cite{Hu2018, Tsinganos2018}.

However, a major disadvantage of the classical ML pipeline is the strong reliance on domain-specific knowledge, needed for feature extraction, feature selection and parameter tuning. The reason why deep learning is appealing for sEMG-based gesture recognition is that it loosens these requirements, replacing feature extraction with feature learning incorporated in the algorithm training.

The general problem of finding good discriminative hand-crafted features is so hard that the quest for better ways to capture the temporal and frequency information of the signal has characterized decades of research. A clear overview of the typical sEMG features identified and used can be found in \cite{Tsinganos2018}, and can be summarized by domain as follows:
\begin{itemize}\itemsep0em
\item time domain: Root Mean Square (RMS), variance, Mean Absolute Value (MAV), Zero Crossings (ZC), Slope Sign Changes (SSC), waveform length, histogram;
\item frequency domain: Short Time Fourier Transform (STFT), cepstral coefficients;
\item time-frequency domain: Marginal Discrete Wavelet Transform (MDWT).
\end{itemize}

Two significant studies on sEMG classification with traditional ML techniques were made by Hugdings in \cite{Hudgins1993} and Englehart and Hudgins \cite{EnglehartHudgins2003}, who studied the classification problem of 4 hand gestures and obtained a classification accuracy higher than 90\% by working with \SI{200}{\ms} segments of 4 channel sEMG signals, extracting 5 time-domain features and feeding them to MLP and LDA classifiers, robustifying the assignment by applying a majority vote window to the predictions. Instead, Castellini et al.\ \cite{Castellini2009} worked on 3 types of grasp motions and achieved a 97.1\% classification accuracy using the RMS value from 7 electrodes as the input to an SVM classifier.

The first successes in the classification of a large number of hand gestures were obtained by Kuzborskij et al.\ \cite{Kuzborskij2012}, using any of the proposed features in both time- and frequency-domain, fed to a SVM classifier with Radial Basis Function (RBF) kernel, and reaching a 70-80\% accuracy on the 52 hand gestures of the 8-channel database Non-Invasive Advanced Prosthetics (NinaPro) presented by Atzori et al.\ in \cite{Atzori2015}. These results where improved by Atzori et al.\ \cite{Atzori2014}, by considering linear combination of features and using a RF classifier resulting in an average accuracy of 75.32\%; and further improved by Gijsberts et al.\ in \cite{Gijsberts2014}, by evaluating different kernel classifiers jointly on EMG and acceleration signals, increasing classification accuracy by 5\%.

\subsection{Deep Learning Revolution}
In recent years, sEMG-based gesture recognition has seen a progressive shift from traditional machine learning to deep learning. Deep Learning (DL) is the class of machine learning algorithms that differ from classical ML approaches in that feature extraction is part of the model definition, therefore obviating the need for hand-crafted features. 

Existing deep learning architectures are mainly based on two kinds of architecture: the Convolutional Neural Network (CNN), successfully deployed for image classification due to its ability to capture spatial (but also temporal) information of the signal, and Recurrent Neural Network, which allows to exploit the sequential nature of the data and has had successes in speech recognition. The work of this master thesis is exclusively focused on the CNN architecture. Although deep algorithms themselves are not new, they are the most computationally demanding, so that deep classifiers gained attention only relatively recently, thanks to increased availability of data and powerful improvements in computing hardware \cite{Goodfellow2016}.

The advantage of deep architectures is their ability to incorporate feature learning: a consistent part of the traditional pipeline can be entrusted to the training of the algorithm, which has enough capacity to learn good feature representations on its own. This mitigates the reliance on rigid combinations of preprocessing steps and precise sets of hopefully discriminative features.

% In (Shim and Lee, 2015) and (Shim et al., 2016), the authors propose a Deep Belief Network (DBN) classifier as a more effective model compared to a shallow MLP network trained with back-propagation. Time-domain features are extracted from segments of 2 channel EMG signals which are used to train the model in a layer-by-layer fashion, either with a greedy approach or using genetic algorithms, achieving an accuracy of 88.59\% and 89.29\% respectively on a set of 5 movements.

The first end-to-end DL architecture was proposed by Park and Lee \cite{ParkLee2016}, who built a CNN-based model for the classification of six common hand movements resulting in a better classification accuracy compared to SVM. Atzori et al.\ \cite{Atzori2016} proposed a simple CNN architecture based on 5 blocks of convolutional and pooling layers to classify the 52 hand gestures from the NinaPro database \cite{Atzori2015}, reaching classification accuracy comparable to those obtained with classical methods, though not higher than the best performance achieved on the same problem using a RF classifier.

Geng et al.\ \cite{Geng2016} and Wei et al.\ \cite{Wei2017} improved the results across various datasets by adding batch normalization and dropout (layer types explained in Subsections \ref{BN} and \ref{dropout}, respectively) to the model architecture, and by using a high-density electrode array, thus benefiting from the setup of High Density sEMG (HD-sEMG). Building the analysis on instantaneous EMG images \cite{Geng2016} achieves a 89.3\% accuracy on a set of 8 movements, going up to 99.0\% when using majority voting over \SI{40}{\ms} signal windows. In \cite{Wei2017} it is shown that for some movements a significant role is played by a small, group of muscles, and therefore a multi-stream CNN architecture is used that divides the inputs into smaller images, to be separately processed by convolutional layers before being merged with fully connected layers, reporting a increase in accuracy by 7.2\% (from 77.8\% to 85\%).

Successful approaches also involve multi-modality data (also called sensor fusion): Du et al.\ \cite{Du2017-2} exploit the auxiliary information of a data glove to add semi-supervised auxiliary tasks (regression on glove signals) to a CNN, reporting improved classification accuracy.

\subsection{State-of-the-art on the variability factors}
Currently, the state-of-the-art of sEMG-based gesture recognition is coping with challenging issues. The sEMG signal is severely affected by many variability factors such as differences between subjects, fatigue, user adaptation and the variability introduced from the re-positioning the electrodes at each session of data collection. These issues limit long-term use and temporal reliability of the devices relying on sEMG analysis.

In the framework of machine learning, these factors of variability can be modeled with the concept of data sources, i.e.\ data subsets coming from different distributions (the concept of multi-source data is defined in Subsection \ref{multisource}). The Multi-source nature of the data makes machine learning on EMG data a challenging problem, where the ambition is to improve the inter-source generalization capability of classifiers. Examples of this are the inter-posture, inter-session and inter-subject scenarios. Up to now, for instance, the performance with Leave-One-Subject-Out cross-validation (LOSOCV) is still much lower than that reached with Within-Subject cross-validation (WSCV) on the main public benchmark databases \cite{Du2017}.

Some studies have already dealt with inter-subject variability, resorting to recalibration techniques
%\cite{Zhai2017}
or model adaptation methods
%\cite{Du2017, CoteAllard2018}
often moving from the technique of batch normalization (explained in Subsection \ref{BN}). The network proposed in \cite{Zhai2017} takes as input downsampled spectrograms (i.e.\ time-frequency representations) of sEMG segments, and improvement is achieved by updating the network weights using the predictions of previous sessions corrected by majority voting. In \cite{Du2017} it is assumed that, while the weights of each layer of the network learn information useful for gesture discrimination, the mean and variance of the batch normalization layers store information related to discrimination between sessions/subjects. Moreover, a variation of batch normalization called Adaptive Batch Normalization (AdaBN) is used in \cite{Li2016}: only the normalization statistics are updated for each subject using a few unlabeled data, improving performance with respect to a model without adaptation.

In \cite{CoteAllard2018} transfer learning techniques are used to exploit inter-subject data learned by a pre-trained source network. In this architecture, for each subject a new network is instantiated with weighted connections to the source network, and predictions for a new subject are based both on previously learned information and subject-specific data. Doing so achieves an accuracy of 98.3\% on 7 movements.

\subsection{State-of-the-art on the Unibo-INAIL dataset}

The state-of-the-art of sEMG-based hand gesture recognition on the Unibo-INAIL dataset is exposed by Milosevic et al.\ in \cite{Milosevic2017}. The dataset itself represents the state of the art with regard to databases able to explore inter-subject, temporal, and postural variability, and is extensively described in Section \ref{dataset}.

Knowledge about the performance of classifiers on the dataset is limited to classical machine learning classifiers, namely Support Vector Machine (SVM), Neural Network (NN) with only one hidden layer, Random Forest (RF) and Linear Discriminant Analysis (LDA). However, these classifiers have been used to explore a large number of data partitions and training strategies, trying to optimize the generalization capability on new arm postures and new days.

The evaluated algorithms have similar recognition performance, higher than 90\% (precise values depending on data partition and training strategy). The Radial Basis Function (RBF)-kernel SVM is at present the one achieving the highest accuracy, both intra-session and inter-session.

With regard to inter-session generalization, the baseline classification accuracy was higher than 90\% for intra-posture and inter-day analysis, suffering a degradation of up to 20\% when testing on data from different postures or days. This is consistent with the inter-session accuracy decline shown for other datasets (e.g.\ the 27\% morning-to-afternoon decline reported for the NinaPro Database 6 by Palermo et al.\ \cite{Palermo2017}). The work showed that this inter-posture and inter-day accuracy decline is mitigated by training with combinations of data from multiple sessions. Moreover, results on temporal variability show a progressive user adaptation trend and indicate that (re)training strategies should prioritize the availability of recent data.

However, despite the variety of strategies and the number of results, the state of the art has two limitations. The first limitation is that, as said above, the state of the art on this dataset is limited to classical machine learning algorithms. The second limitation concerns data preprocessing: all the algorithms are applied on singled out, instantaneous signal values (i.e.\ 4-dimensional data points whose only dimension is channel number), so that the sequential nature of the EMG signal is not exploited at all.  This master thesis addresses both these limitations.

%...

%...

% GRAFICI E/O TABELLE

% Grafici e tabelle per mostrare meglio le strategie e i risultati.

% % % % % % % % % % % % % % % % % % % % % % % % % % % % % % % % % % % % % % % % % % % % % % % % % % % % % % % % % % % % % % % % % % % % % % % % % % % % % % % % % % % % % % % % % % % % % % % % % % % % % % % % % % % % % % % % % % % % % % % % % % % % % % % % % % % % % % % % % % % % % % % % % % % % % % % % % % % % % % % % % % % % % % % % % % % % % % % % %
\newpage
\chapter{Deep Learning and Convolutional Neural Networks}\label{DL}

Deep learning is the sub-field of machine learning which deals with deep neural networks, i.e.\ neural networks having more than one hidden layer. Deep learning is currently the fundamental approach for the development of artificial intelligence. One of the most important deep neural network architectures is the convolutional neural network, which is the model used in this work.

This chapter is structured as follows:
\begin{itemize}\itemsep0em
\item[$\circ$] Section \ref{NN_DL} introduces neural networks and deep learning, contextualizing them with respect to machine learning and artificial intelligence;
\item[$\circ$] Section \ref{train} provides the fundamentals of deep networks training, and illustrates some specific techniques and settings used in this work;
\item[$\circ$] Section \ref{CNN} introduces convolutional neural networks, explaining the essential layer types (i.e.\ convolutional and fully connected) and also introducing the other particular layers used in this work.
\end{itemize}

\section{Neural Networks and Deep Learning}\label{NN_DL}

Deep learning (DL) is the branch of Machine Learning (ML) which is nowadays the base of the research and applications of Artificial Intelligence (AI) \cite{LeCun2015}, and in the last decade has produced vast advances in many fields such as speech recognition \cite{Deng2013} and image recognition \cite{Krizhevsky2012}, cancer detection \cite{Esteva2017}, self-driving vehicles \cite{Chen2015}, and playing complex games \cite{Silver2016}. In some applications, DL models already exceed human performance.

The superior performance of DL models resides in the ability to incorporate automated data-driven feature extraction selection, i.e.\ the ability to extract high-level features from raw data by using statistical learning on large datasets. This approach produces effective representations of the inputs space, based on powerful discriminative features, in a different way from the earlier approaches based on non-deep ML, which are built on hand-crafted features or rules designed by expert researchers. % However, increased performance comes at the cost of higher computational complexity. This is due to the fact that effective deep representations can be learned only by complex models

\begin{figure}[H]
  \centering
  \begin{subfigure}[b]{0.5\textwidth}
  \includegraphics[width=\textwidth]{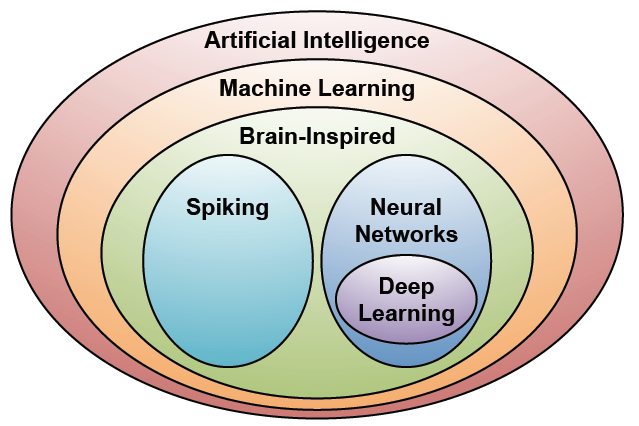}
  \end{subfigure}
  \caption{Deep Learning contextualized with respect to Artificial Intelligence and Machine Learning. Image from \cite{Sze2017}.}
  \label{img_DL_sets}
\end{figure}

It is important to contextualize clearly the relationship DL has with AI and ML, as illustrated in Figure \ref{img_DL_sets}. With respect to the broad domain of AI, ML can be considered a vast sub-domain. Inside ML, is the area referred to as brain-inspired computation, whose interest is the developing of programs or algorithms whose basic functionality is inspired by natural brains and aims to emulate some aspects of how we currently understand the brain works \cite{Sze2017}. Brain-inspired computing is divided into to main branches: spiking computing and DL. Spiking computing, which is beyond the scope of this work, takes inspiration from the fact that communication between neurons happens via spike-like pulses, with information not simply coded in spike's amplitude, but also in the timing of the spikes. In contrast, the branch of brain-inspired computing relevant here is Neural Networks (NNs), which are a well-known ML algorithm.

\begin{figure}[H]
  \centering
  \begin{subfigure}[b]{0.5\textwidth}
  \includegraphics[width=\textwidth]{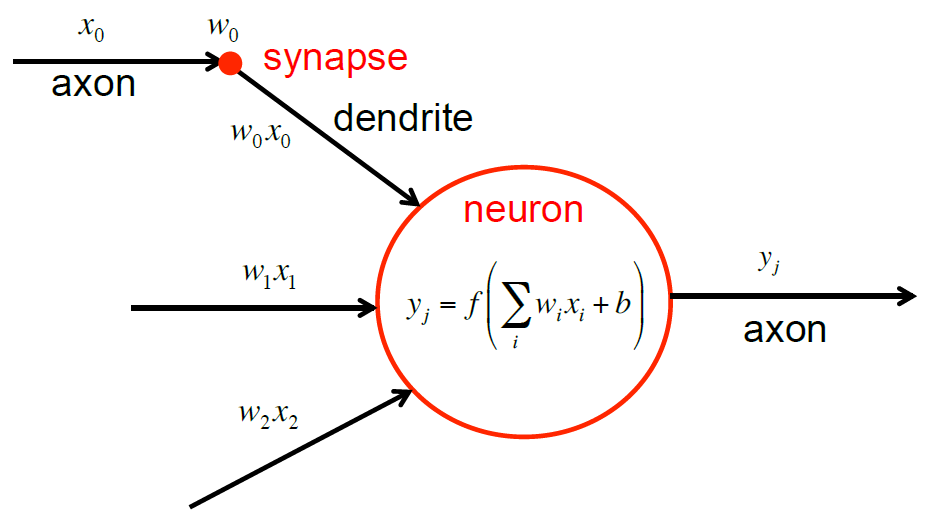}
  \end{subfigure}
  \caption{Model of the computations performed by a brain neuron: $x_i$ are the input activations, $w_i$ are the weights of the weighted sum, $f(\cdot)$ is the non-linear function, and $b$ is the bias term. Image from \cite{Li2018}.}
  \label{img_DL_neuron}
\end{figure}

Neural networks are inspired by the fact that the signal processing performed by a neuron can be modeled as a weighted sum of the input activations, followed by a non-linear function with threshold and bias, which generates the neuron's output signal \cite{Li2018} as illustrated in Figure \ref{img_DL_neuron}. By analogy, neural networks are built assembling units (also referred to as neurons) which apply a non-linear function to the weighted sum of the input values they receive.

\begin{figure}[H]
  \centering
  \begin{subfigure}[b]{0.80\textwidth}
  \includegraphics[width=\textwidth]{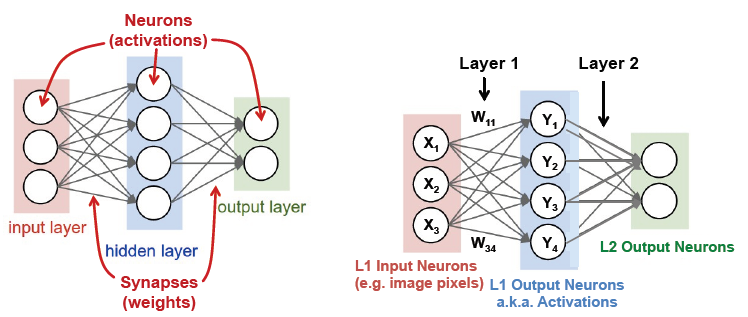}
  \end{subfigure}
  \caption{Scheme of a simple neural network's structure and terminology. Left: neurons and synapses. Right: weighted sum computation for each layer (bias term $b$ omitted for simplicity). Image from \cite{Sze2017}.}
  \label{img_DL_NNs}
\end{figure}

Neural Networks, like the computations they execute, are structured as layers of units, and an example scheme is shown in Figure \ref{img_DL_NNs}. The neurons belonging to the input layer receive the raw input values, and propagate them to the neurons of the middle layer, called the hidden layer. The weighted sums computed by one or more hidden layers ultimately reach the output layer, whose units compute the final output of the network. Taking the brain-inspired terminology further, the neuron outputs are referred to as activations, and the weights are sometimes called synapses. In addition to the network structure, Figure \ref{img_DL_NNs} also shows the computations made at each layer, which follow the formula
\begin{align}
y_j &= f\left(\sum_{i=1}^3 w_{ij} x_i + b_j\right),
\end{align}
where $x_j$ are the input activations, $w_{ij}$ are the weights, $b_j$ are the bias terms, $y_j$ are the output activations, and $f(\cdot)$ is the non-linear activation function.

Deep learning is the field of Neural Networks which focuses on Deep Neural Networks (DNNs), which are networks having more than one hidden layer, or, equivalently, more than three layers in total (typically, from five to more than a thousand \cite{Sze2017}). DNNs have the ability to learn high-level features with more complexity and abstraction than shallow (i.e.\ non-deep) neural networks. For instance, in image processing, pixels of an image are fed to the first layer of a DNN, whose outputs can be interpreted as representing the presence of low-level features in the image, such as lines and edges. Subsequent layers progressively combine these features, eventually yielding a measure associated to the presence of higher level features (e.g.\ edges are combined into shapes, then into sets of shapes). As final output, the network produces an estimate of the probability that the highest-level features comprise a particular object or scene. This paradigm is referred to as deep feature hierarchy, and is what gives the DNNs the ability to obtain superior performance.

\section{Training deep networks: gradient descent and back-propagation}\label{train}

In a DNN, like in machine learning algorithms in general, there is a basic program that does not change while learning to a desired task. For DNNs, the basic program is the structure of the functions implemented by the layers, and learning consists in determining the value of the network's weights and biases, through an optimization called training. Once trained, the network can execute its task by computing the output using the optimized weights and biases. Running the trained networks to evaluate inputs is referred to as inference.

The training scenario relevant in this work is the supervised training for a classification task. In classification tasks, trained DNNs receive input data (e.g.\ the pixels of an image) and return vector of scores, one for each class. The highest-score class is the one the network estimates as most probable. The main goal of DNN training is to optimize the weights and the bias values so as to maximize the score of the correct class and minimize the scores of the incorrect classes. In supervised learning, the correct class of the data the network is trained on is known. The dissimilarity between the ideal correct scores and the scores computed by the DNN (based on its current weights and biases) is called the loss $L$, which is the objective function of the training, i.e. function to minimize.

A widely used algorithm for weight optimization is a hill-climbing iterative optimization procedure called gradient descent. In gradient descent, at each iteration $t$ the weights $w_{ij}$ is updated by subtracting a multiple of the gradient of $L$ with respect to the weights. Element-wise:
\begin{align}
w_{ij}^{(t+1)} = w_{ij}^{(t)} - \alpha \dfrac{\partial L}{\partial w_{ij}}
\end{align}
where the multiplication factor $\alpha$ is called learning rate. Iterating reduces $L$.

In contrast with weights and biases, which are referred to as model parameters since they are the arguments of the objective function to minimize, the learning rate $\alpha$ is not involved in differentiation and hence is a training hyper-parameter, i.e.\ a constant that regulates the optimization without being affected by training. As the hyper-parameters describing the structure or training of any machine learning model, $\alpha$ can be tuned to the optimal value through cross-validation (which can be computationally expensive) or through a shorter preliminary analysis (as done in this work), with a further speed-up commonly yielded by following heuristics instead of grid search. The optimal $\alpha$ is typically small (some orders of magnitude below unity), but the its order of magnitude is strongly dependent on data, task, and model architecture. Hence the search must cover different orders of magnitude.

Finer techniques for tuning the learning rate are scheduling and recourse to per-parameter learning rates. (1) With scheduling, $\alpha$ is reduced over the iterations, typically with a stepwise or exponential decay, in order to adapt the tuning to different moments of the gradient descent. (2) On the other hand, introducing per-parameter learning rates allows to perform a customized tuning on different parameters or groups of parameters. Both techniques come at the cost of an increased number of combination to explore.

\begin{figure}[H]
  \centering
  \begin{subfigure}[b]{0.67\textwidth}
  \includegraphics[width=\textwidth]{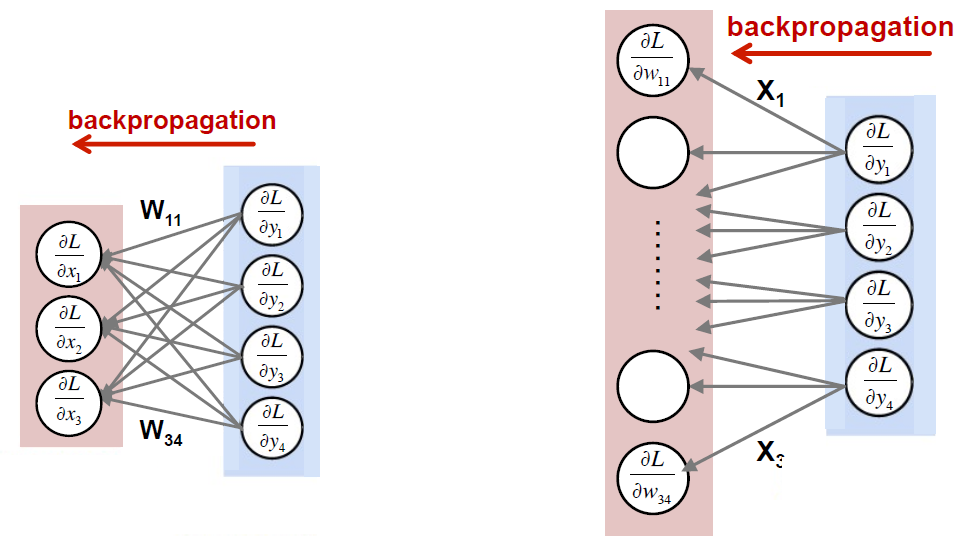}
  \end{subfigure}
  \caption{Example of back-propagation through a neural network: (1) (left) computation of the gradient of the loss relative to the layer inputs; (2) (right) computation of the gradient of the loss relative to the weights. Image from \cite{Sze2017}.}
  \label{img_DL_backprop}
\end{figure}

An efficient procedure for computing the partial derivatives of the loss is back-propagation, which derives from the
chain rule of calculus and operates by passing values backwards through the network to compute how the loss is affected by each weight. Computation using back-propagation, illustrated in Figure \ref{img_DL_backprop}, requires some steps used also for inference. To back-propagate through each layer, one has to: (1) compute the gradient of $L$ relative to the weights from the layer inputs (i.e., the forward activations) and the gradient of $L$ relative to the layer outputs; (2) compute the gradient of $L$ relative to the layer inputs from the layer weights and the gradients of $L$ relative to the layer outputs. It is worth to note that back-propagation requires intermediate activations to be preserved for the backward computation, so that training has increased storage requirements compared to inference.

%Second, due to the gradients use for hill-climbing, the precision requirement for training is generally higher than inference. Thus many of the reduced precision techniques discussed in ... are limited to inference only.
% A variety of techniques are used to improve the efficiency and robustness of training. For example, often the loss from multiple sets of input data, i.e., a batch, are collected before a single pass of weight update is performed; this helps to speed up and stabilize the training process.

Another popular training method, orthogonal to the techniques exposed so on, is fine-tuning. In fine tuning, the weights of a trained network are available, and are used as a starting point for the iterative optimization. This practice results in faster training than using random initialization. Moreover, the scenario in which the weights are adjusted for a new dataset is the sub-field of machine learning defined transfer learning.

More specific training settings and techniques used in this work are explained in the next subsections.

\subsection{Cross-entropy loss}\label{crossentropy}

When training neural networks, a convenient choice for the objective function (adopted also in this work) is the cross-entropy function. Originating in the field of probability theory and information theory, the main virtue of cross-entropy is that it leverages the soft assignments produced by the network, interpreting them as probabilities. Given a multiclass classification task on $C$ classes, the loss for the observation $x_i$ is computed by summing the losses due to each soft assignment $\hat{y}_{ic}$ with respect to the true label $y_{ic}$ (i.e.\ the one-hot encoding):
\begin{align}
\textrm{loss}(x_i) = - \sum\limits_{c = 1}^{C} y_{ic} \log \hat{y}_{ic},
\end{align}
%\ begin{align}
%\textrm{loss}(x,\textrm{class})
%&= - \log \left(\dfrac{\exp(x[\textrm{class}])}{\sum_j\exp(x[j])}\right)
%&= - x[\textrm{class}] + \log\sum_j\exp{x[j]},
%\ end{align}
with arbitrary choice for the base; e.g., taking the logarithm in base 2 yields a result in bits, whereas choosing base $e$ yields a result in nats. The overall loss $L$ is then computed by averaging over all the items to classify. When working with an unbalanced dataset, it can be useful to weight the elements by class when taking the average.

\subsection{Stochastic gradient descent with mini-batches}\label{SGD}

Stochastic Gradient Descent (SGD) is a variation of gradient descent which helps avoiding local minima during training. This technique does so introducing randomness in the optimization process by randomly partitioning the training set, referred to as the batch, into $B$ equal-sized subsets called mini-batches. Then, gradient descent is performed using a single mini-batch per iteration, cycling over all the mini-batches. The sequence of $B$ iterations that processes all the mini-batches exactly once is referred to as an epoch. Thus, during an epoch, all the data contribute to the optimization to the same amount. At the end of each epoch, the split into mini-batches is re-randomized, in order to prevent systematic bias generated by a particular order of comparison of the data.

It is important to note that the avoidance of local minima provided by SGD comes at the cost of introducing a new training hyper-parameter, the batch size $b$ (or, equivalently, the number of batches $B$).
% Common values used for $b$ are powers of 2 less than or equal to 1024.

Since some sources reserve the name SGD for the extreme case $b = 1$, in the remainder of this work the term \textit{SGD with mini-batches} will be used to avoid confusion.

\subsection{L$\mathbf{_2}$ regularization}\label{L2regularization}

The technique of L$_2$ regularization takes its name from the L$_2$-norm and is a method to counter overfitting. It consists in adding to the loss a multiple of the squared L$_2$-norm of the vector of all the parameters, and using the new formula $L_\text{reg}$ as objective function. In formulas, for a deep neural network:
\begin{align}
L_\text{reg} &= L_0 + L_{\text{L}_2} = \\
&= L_0 + \lambda_{\text{L}_2} \sum\limits_{l = 2}^{N_\text{layers}}
\sum\limits_{i = 1}^{n_{l - 1}} \sum\limits_{j = 1}^{n_l} |w_{ij}^{(l)}|^2,
\end{align}
where $L_0$ is the non-regularized loss, $L_{\text{L}_2}$ is the regularization term, $N_\text{layers}$ is the number  of layers, $n_l$ is the number of neurons in layer $l$, and $w_{ij}^{(l)}$ is the weight connecting neuron $i$ of layer $l - 1$ to neuron $j$ of layer $l$. The factor $\lambda_{\text{L}_2}$ is the training hyper-parameter regulating the amount of regularization enforced. Too low $\lambda_{\text{L}_2}$ has no effect, and too high $\lambda_{\text{L}_2}$ incurs underfitting. The optimal value can be determined via preliminary analysis or cross-validation, covering different orders of magnitude.

%Regularization aim, $L_2$, $L_1$ and sparse representations, $L_2 \leftrightarrow \textrm{Gaussian priors on \textit{w}'s}$.

\section{Convolutional Neural Networks}\label{CNN}

Deep networks have a vast variety of architectures and sizes, continuously evolving to increase performance. Convolutional neural networks are a successful class of architectures, which can be introduced only after surveying the strategies used to progressively reduce the storage and computation required by layers: sparsity, structured sparsity, weight sharing, and convolution.

\begin{figure}[H]
  \centering
  \begin{subfigure}[b]{0.5\textwidth}
  \includegraphics[width=\textwidth]{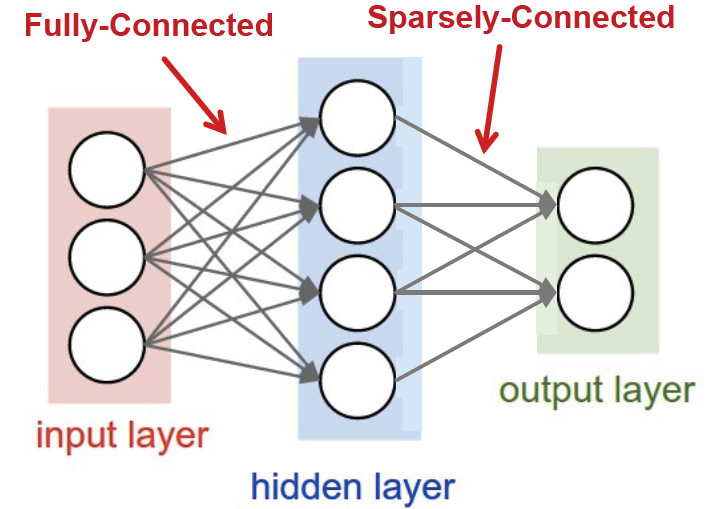}
  \end{subfigure}
  \caption{Fully conncted layer versus sparsely connecetd layer. Image from \cite{Sze2017}.}
  \label{img_DL_FCSC}
\end{figure}

Deep networks can be entirely composed of fully-connected (FC) layers, shown in Figure \ref{img_DL_FCSC}, and in this case they are defined Multi-Layer Perceptrons (MLP). In a FC layer, all outputs units are connected to all inputs units, so that all output activations are computed with a weighted sum of all input activations. Although the FC configuration requires significant computation and storage, in many situations it is possible to zero some weights (thus removing the relative connections) without impacting performance. The resulting layer, also shown in Figure \ref{img_DL_FCSC}, is called sparsely connected layer.

\begin{figure}[H]
  \centering
  \begin{subfigure}[b]{0.75\textwidth}
  \includegraphics[width=\textwidth]{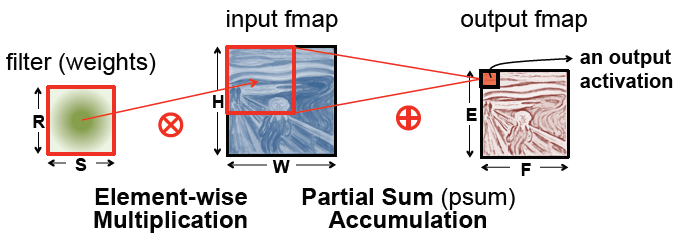}
  \end{subfigure}
  \caption{2d convolution in traditional image processing. Image from \cite{Sze2017}.}
  \label{img_DL_convolution}
\end{figure}

Opposed to generic sparsity, structured sparsity is the configuration in which each output is only a function
of a fixed-size window of inputs. Even further efficiency is gained when the computation of every output employs the same set of weights. This configuration is known as weight sharing, and strongly reduces the storage requirements for weights. A particular case of weight sharing arises when the computation is structured as a convolution, as shown in Figure \ref{img_DL_convolution}: the weighted sum for each output activation is computed using only a narrow neighborhood of input activations (by zeroing the weights outside the neighborhood), and every output shares the same set of weights (i.e., the filter is space invariant). This gives rise to convolutional (CONV) layers, which are the characteristic building block of convolutional neural networks.

\begin{figure}[H]
  \centering
  \begin{subfigure}[b]{0.75\textwidth}
  \includegraphics[width=\textwidth]{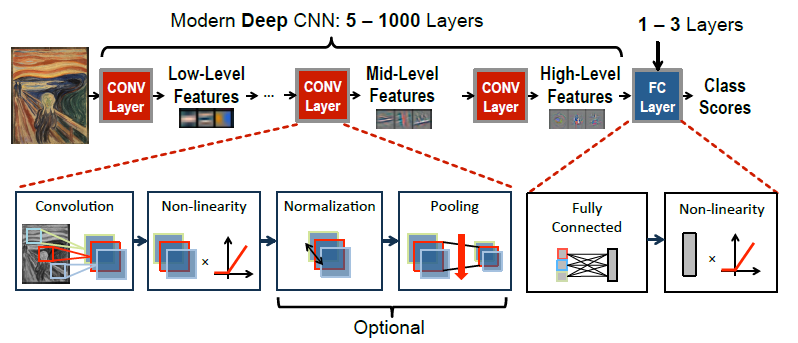}
  \end{subfigure}
  \caption{A convolutional neural network. Image from \cite{Sze2017}.}
  \label{img_DL_CNN}
\end{figure}

Convolutional Neural Networks (CNNs) are a successful deep network architecture, composed of multiple CONV layers, as shown in Figure \ref{img_DL_CNN}. In CNNs, each layer generates a progressively higher-level representation of the input data, referred to as feature map (fmap), extracting the essential information for the network's task. Modern CNNs have attained superior performance by implementing a very deep hierarchy of layers. CNN are widely used in a variety of applications including image understanding \cite{Krizhevsky2012}, speech recognition \cite{Sainath2013}, robotics \cite{Levine2016} and game play \cite{Silver2016}. In this work, CNNs are applied to the task of classifying time windows of a 4-channel signal.

CNNs fall into the category of feed-forward networks. In a feed-forward network, all computations are executed as a sequence of operations taking place from one layer to the next one. A feed-forward network has therefore no memory, and the output for a given input is always identical irrespective of the history of the inputs fed previously.

\begin{figure}[H]
  \centering
  \begin{subfigure}[b]{0.75\textwidth}
  \includegraphics[width=\textwidth]{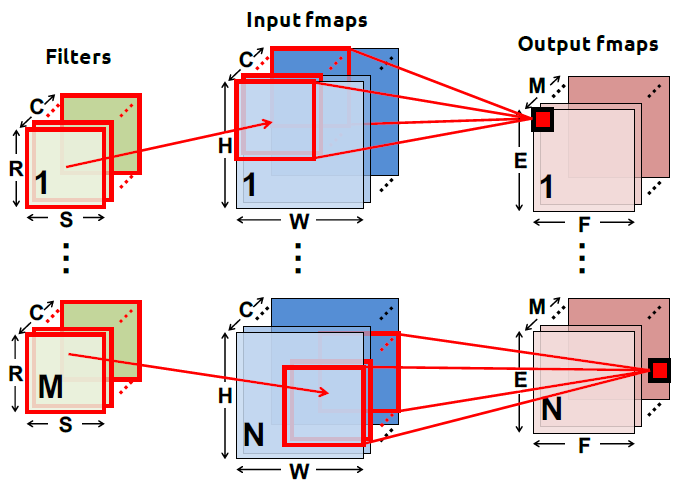}
  \end{subfigure}
  \caption{High-dimensional convolutions in CNNs. Image from \cite{Sze2017}.}
  \label{img_DL_hdconvCNN}
\end{figure}

\newpage
Each CONV layer is mainly constituted by high-dimensional convolutions, as shown in Figure \ref{img_DL_hdconvCNN}. In this computation, the input activations of a layer have the structure of a set of 2d input feature maps (ifmaps), each referred to as a channel. Each channel is convolved with a distinct 2d filter from the stack of filters, one for each channel. The stack of 2d filters being a 3d structure, it is sometimes collectively called a 3d filter. The results of the convolutions at each point are summed across channels, and a 1d bias is optionally \cite{He2016} added to the results. The result of this computation are the output activations constituting one channel of the output feature map (ofmap). Additional output channels can be created by applying additional 3d filters on the same input.
% Multiple input feature maps can be processed together as a batch to potentially improve reuse of the filter weights.

With the notation for shape parameters defined in Table \ref{table_shapeparameters}, the computation executed by a CONV layer is described by the formula
\begin{align}
\mathbf{O}[z][u][x][y] = \mathbf{B}[u] +
\sum\limits_{k = 0}^{C - 1}
\sum\limits_{i = 0}^{S - 1}
\sum\limits_{j = 0}^{R - 1}
\mathbf{I}[z][k][U x + i][U y + j] \times \mathbf{W}[u][k][i][j],
\end{align}
with
\begin{align}
0 \leq z < N, \quad 0 \leq u < M, &\quad 0 \leq x < F, \quad 0 \leq y < E,\\
E = \dfrac{H - R + U}{U}, &\quad F = \dfrac{W - S + U}{U},
\end{align}
where \textbf{O}, \textbf{I}, \textbf{W} and \textbf{B} are the matrices of ofmaps, ifmaps, filters and biases, respectively, $U$ is a fixed stride size. A graphical representation of this computation is shown in Figure \ref{} (where biases are omitted for simplicity).

To align the CNN terminology with the general DNN terminology, is it worth remarking that
\begin{itemize}\itemsep0em
\item filters are composed of weights (corresponding to synapses in nature);
\item input and output feature maps (ifmaps, ofmaps) are composed of activations of inputs and output neurons.
\end{itemize}

\begin{table}[H]
\centering
\begin{tabular}{@{}cc@{}}
\toprule
\textbf{Shape parameter} & \textbf{Meaning}                          \\ \midrule
$N$               & batch-size of 3d maps                            \\
$M$               & number of 3d filters / number of batch channels  \\
$C$               & number of ifmap/filter channels                  \\
$H/W$             & ifmap plane heigth/width                         \\
$R/S$             & filter plane heigth/width ($= H$ or $W$ in FC)   \\
$E/F$             & ofmap plane heigth/width ($= 1$ in FC)           \\ \bottomrule
\end{tabular}
\caption{Shape parameters of a CONV/FC layer.}
\label{table_shapeparameters}
\end{table}

In the CNN used in this work, the CONV layers are CONV-1d layers, i.e.\ layers performing a 1-dimensional convolution. Though less common, CONV-1d layers follow the same principles as CONV-2d layers, but work with inputs having 1 dimension (plus channel number). The choice to use CONV-1d layers was made in order to act on the time dimension of the signal, while using the 4 sEMG channels as CNN input channels. In addition to CONV layers and FC layers, the CNN implemented in this work contains other elements, namely the rectified linear unit, batch normalization and dropout, whose mechanisms are explained in the following two subsections. These three kinds of intervention on activations are sometimes conceptualized as layers.

\subsection{Batch-normalization}\label{BN}

Controlling the input distribution across layers can speed up training and improve accuracy. Accordingly, the distribution of a layer's input activations (described by its mean $\mu$ and standard deviation $\sigma$) can be standardized to zero mean and unit standard deviation. Batch Normalization (BN) \cite{Ioffe2015} is the technique in which the standardized activations are further scaled and shifted, undergoing the transformation
\begin{align}
y = \dfrac{x - \mu}{\sqrt{\sigma^2 + \varepsilon}}\gamma + \beta,
\end{align}
where the parameters $\gamma$ and $\beta$ are learned from training, and $\varepsilon$ is a small constant used to avoid numerical problems. Batch normalization is mostly applied between the CONV or FC layer and the non-linear activation function, and is usually turned off after training.

\subsection{ReLU activation function}\label{ReLU}

\begin{figure}[H]
  \centering
  \begin{subfigure}[b]{0.67\textwidth}
  \includegraphics[width=\textwidth]{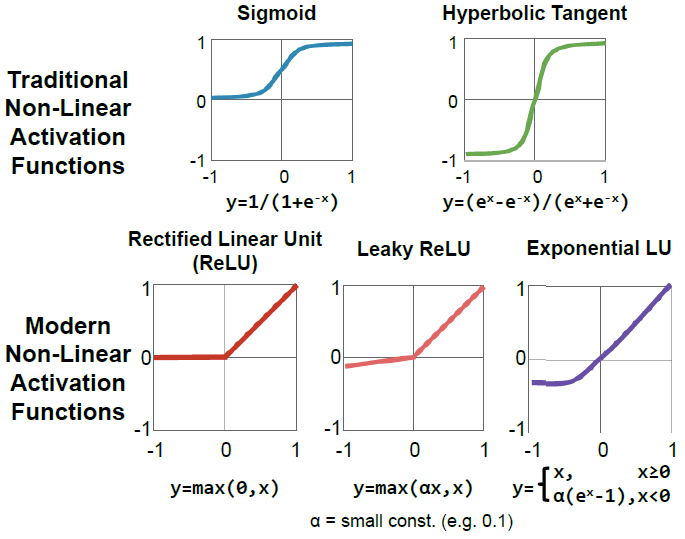}
  \end{subfigure}
  \caption{The most common non-linear activation functions. Image from \cite{Jia2014}.}
  \label{img_DL_nlaf}
\end{figure}

Non-linear activation functions are typically applied after each CONV or FC layer. The most common functions used to introduce non-linearity into a DNN are shown in Figure \ref{img_DL_nlaf}. Historically, the sigmoid and the hyperbolic tangent are the most conventional, while the Rectified Linear Unit (ReLU) has become common in the last years due to its simplicity and its ability to make training faster \cite{Nair2010}. The leaky ReLU, parametric ReLU, and exponential LU are variations of the ReLU explored for increased accuracy. ReLU is mostly applied after the CONV or FC layer or after batch normalization (if present).

\subsection{Dropout}\label{dropout}

Dropout is a technique to improve accuracy by reducing overfitting \cite{Hinton2012, Srivastava2014}. It works by randomly dropping units (and their connections) from the network during training. This prevents the phenomenon of units co-adaptation, forcing each neuron to learn a feature helpful for computing the correct output.
%During training, dropout samples from an exponential number of different “thinned” networks. At test time, it is easy to approximate the effect of averaging the predictions of all these thinned networks by simply using a single unthinned network that has smaller weights.
In particular, the random dropout of the units of the interested layer during training is regulated by the drop probability $p$, for which a common value is $p = 0.5$. On every forward pass, each unit is zeroed out independently and randomly, drawing from the Bernoulli distribution parameterized by $p$. Moreover, the outputs are multiplied by $\frac{1}{1-p}$. Dropout is active only during training: during inference, the dropout layer is disabled (or, equivalently, the drop probability $p$ is set to 0). Dropout is mostly applied immediately before the CONV or FC layer.

% % % % % % % % % % % % % % % % % % % % % % % % % % % % % % % % % % % % % % % % % % % % % % % % % % % % % % % % % % % % % % % % % % % % % % % % % % % % % % % % % % % % % % % % % % % % % % % % % % % % % % % % % % % % % % % % % % % % % % % % % % % % % % % % % % % % % % % % % % % % % % % % % % % % % % % % % % % % % % % % % % % % % % % % % % % % % % % % %
\newpage
\chapter{Materials and Methods}\label{matmet}

This chapter explains all the materials and methods used in this master thesis. It is structured as follows:
\begin{itemize}\itemsep0em
\item[$\circ$] Section \ref{dataset} exhaustively describes the Unibo-INAIL dataset;
\item[$\circ$] Section \ref{pipeline} details the implemented machine learning pipeline, preprocessing, CNN architecture and training settings;
\item[$\circ$] Section \ref{trainingstrategies} defines the concept, essential in this work, of training strategy based on training set composition, and illustrates the training strategies used.
\end{itemize}

\section{Unibo-INAIL dataset}\label{dataset}

The work of this master thesis is entirely focused on the Unibo-INAIL dataset. The Unibo-INAIL dataset is a surface electromyography (sEMG) dataset realized to explore the impact of arm posture and temporal variability (either alone or combined) on sEMG-based hand gesture recognition. The dataset was presented by Milosevic et al.\ in \cite{Milosevic2017}, and was built on the preliminary analysis carried out by Benatti et al.\ in \cite{Benatti2014}. This master thesis evaluates the performance of Convolutional Neural Networks (CNNs) trained on the dataset with all the training strategies described in \cite{Milosevic2017}, where the performance of classical machine learning algorithms is reported.

The data were acquired from 7 able-bodied (i.e.\ non-amputee) subjects performing 6 discrete hand gestures in 4 arm postures, repeated for 8 days, thus probing a total of 224 different data sources, each identified by three discrete indexes (subject, day and arm posture) and containing 6 classes, i.e.\ five hand gestures plus the rest position. This acquisition protocol, in addition to investigating the signal patterns associated to the 6 classes, allows to characterize the following sources of variability affecting signal and patterns:
\begin{itemize}\itemsep0em
\item inter-posture variability: for each subject individually, keeping sensors on, different arm postures and wrist orientations cause variability in signal and patterns, due to differences in muscle activity and muscle position (since sensors are only fixed with respect to the skin);
\item inter-day variability: for each subject individually, temporal variability of signal and patterns are mainly due to two factors:
\begin{itemize}
\item sensor placement: from day to day, the sensors are removed and repositioned, causing differences in signal and patterns, due to the change of the relative position of the electrodes with respect to the muscles;
\item user adaptation: user adaptation (explained in Subsection \ref{useradaptation}) is the transient observed when the inter-day differences in gesture execution decrease over time, due to the fact that new users adapt to the repetitive exercise during the first days;
\end{itemize}
\item inter-subject variability: signal and patterns are influenced by the anatomical variability between subjects (even if all able-bodied).
\end{itemize}

For the research on sEMG-based hand gesture recognition for reliable Human-Machine Interfaces (HMIs), the Unibo-INAIL dataset is a valuable ground for two reasons. The first reason is that the acquisition setup is based on commercial sensors, chosen so as to make the setup is easily repeatable and thus suitable for integrated HMI controllers \cite{Benatti2017}. The second reason is that the Unibo-INAIL dataset is the first public sEMG dataset to date to include both arm-postural and session variability, providing a realistic scenario for evaluating classification algorithms and training approaches.

\subsection{Unibo-INAIL collaboration and motivation for the dataset}

The Unibo-INAIL dataset is the results of a research project funded by the Istituto Nazionale per l'Assicurazione contro gli Infortuni sul Lavoro (INAIL), in which the University of Bologna (Unibo) was designated for assessing the feasibility of real-time control of poly-articulated hand prostheses by means of pattern recognition algorithms implemented on a microcontroller. The project was inspired by a previous study by Castellini et al.\ \cite{Castellini2009} on intuitive prosthesis control, which demonstrated that SVMs can recognize different muscle activation patterns with high precision. In particular, the SVMs classify gestures up to a precision of 95\% and approximate the forces with an error of as little as 7\% of the signal range, sample-by-sample at \SI{25}{\Hz}.

The first part of the project aimed to assess how accurate the recognition can be on diverse data produced in different conditions (i.e.\ intra-session scenario), and to verify whether the system was stable on different sessions (i.e.\ inter-session generalization). Although in prosthetics sensor (re)positioning is less relevant, since sensors are fixed to the prosthesis and thus much less mobile, this variability factor was included, planning future developments.

The second branch of the project involved the validation of the algorithms in real-time control scenario. The controller was implemented on a microcontroller, in a system in which the real-time, fresh data were acquired with the same embedded setup used for the first part of the project, in order to reproduce the previous system. The algorithm implementation was also made identical by using the open source machine learning library LIBSVM, which is implemented for both Matlab and C \cite{LIBSVM}.

This master thesis continues the first part of the project, aiming to expand the results obtained for classical machine learning algorithms (SVM, shallow NN, RF and LDA) applied on instantaneous 4-channel signal values. This work extends the analysis to deep CNNs applied on \SI{150}{\ms} time windows of the 4-channel signal.

%Il dataset è frutto di un progetto di ricerca finanziato da INAIL, che ha incaricato Unibo di verificare la fattibilità di una protesi di mano poliarticolata, controllata da algoritmi di pattern recognition, implementati in real-time su microcontrollore. Il lavoro è stato ispirato da un precedente studio (C. Castellini, E. Gruppioni, A. Davalli, G. Sandini, Fine detection of grasp force and posture by amputees via surface electromyography, \textit{Journal of Physiology Paris}, 103:255–262, 2009. Elsevier.) nel quale si dimostrava che l'SVM è in grado di riconoscere con ottima accuratezza diversi pattern di attivazione muscolare per il controllo "intuitivo" di una protesi. Noi siamo partiti da qua.

%La prima parte del lavoro è stata di capire quanto era buono il riconoscimento su una casistica abbastanza vasta, e di vedere se il sistema era stabile su diverse sessioni (come hai visto, non molto). Questo in realta per una protesi è meno rilevante, dato che i sensori vengono di fatto fissati all'aggancio della protesi, quindi sono molto meno "mobili", comunque lo studio lo abbiamo fatto ugualmente anche prevedendo sviluppi futuri.

%Per acquisire i dati abbiamo usato lo stesso setup embedded che poi è stato utilizzato per implementare su micrcontrollore l'algoritmo ed il controller. Questo è stato fatto in modo da avere sempre lo stesso sistema per lo studio offline e per quello online, in modo da minimizzare i possibili problemi. Anche l'implementazione dell'algoritmo era la stessa , dato che abbiamo usato libSVM che è implementata sia per matlab che in C, ed è opensource.

\subsection{Outline of acquisition setup and experimental protocol}\label{acqsetup}
\begin{figure}[H]
  \centering
  \begin{subfigure}[b]{1\textwidth}
  \includegraphics[width=\textwidth]{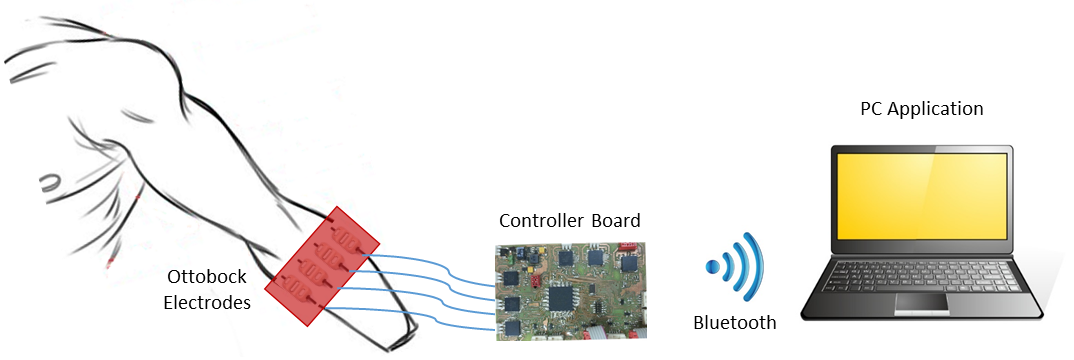}
  \end{subfigure}
  \caption{Acquisition setup of the Unibo-INAIL dataset. Image from \cite{Milosevic2017}.}
  \label{img_dataset_setup}
\end{figure}

The acquisition setup, shown in Figure \ref{img_dataset_setup}, was designed to be reliable and repeatable, with characteristics typical of prosthetic applications. However, since all the data are collected from able-bodied (i.e.\ non-amputee) subjects, the dataset is useful for any HMI application.

\begin{figure}[H]
  \centering
  \begin{subfigure}[b]{0.48\textwidth}
  \includegraphics[width=\textwidth]{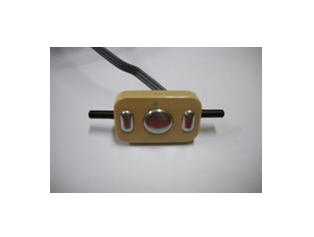}
  \caption{Ottobock sensor used to produce the Unibo-INAIL dataset. Image from \cite{Benatti2014}.}
  \label{img_dataset_ottobock}
  \end{subfigure}
  \hfill
  \begin{subfigure}[b]{0.48\textwidth}
  \includegraphics[width=\textwidth]{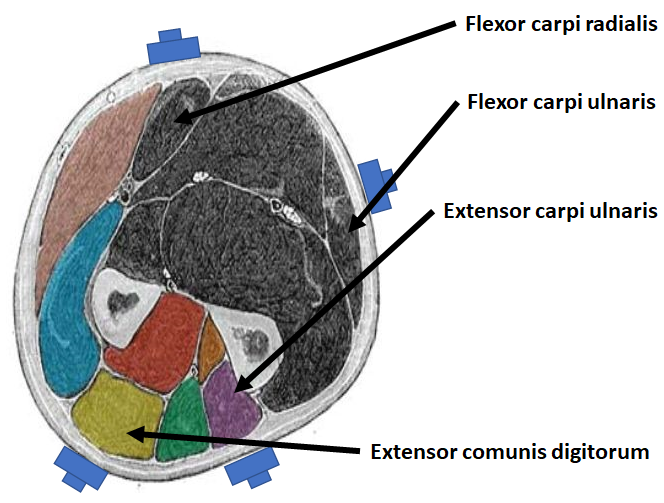}
  \caption{Forearm muscle cross-section and sensors placement. Image from \cite{Milosevic2017}.}
  \label{img_dataset_muscles}
  \end{subfigure}
  \caption{}
  \label{}
\end{figure}

The acquisition setup is based on the Ottobock 13E200 pre-amplified single-ended sEMG electrode (Figure \ref{img_dataset_ottobock}), which is a commercial sensor. It amplifies and integrates the raw EMG signal to reach an output span of $0-3.3\textrm{V}$. %, suitable for the single-ended interface of the ADC of an embedded microcontroller.
The sensors have bandwidth spanning $90-450\textrm{Hz}$ and integrate an analog notch filter to remove the noise due to Power Line Interference (PLI), i.e.\ the capacitive coupling between the subject and the surrounding electrical devices and power grid (detailed in Section \ref{sEMG}). The output analog signals were acquired with a custom embedded board based on a microcontroller equipped with an internal 16-bit ADC. The digitalized signals were streamed via Bluetooth to a laptop, for storage and off-line data analysis.

The subjects involved were able-bodied (i.e.\ non amputee) males, 29.5 $\pm$ 12.2 years. During the acquisition the subjects worn an elastic armband with 4 Ottobock sensors placed on the forearm muscles involved in the selected movements (i.e. \textit{extensor carpi ulnaris}, \textit{extensor communis digitorum}, \textit{flexor carpi radialis} and \textit{flexor carpi ulnaris}) as shown in Figure \ref{img_dataset_muscles}. Sensors were placed on the proximal third at \SI{30}{\mm} respectively on the left and on the right side of two axial lines ideally traced on the forearm. Each acquisition consisted in 10 repetitions of each hand gesture, with 3 second contractions interleaved by 3 seconds of muscular relaxation to be later labeled as rest gesture. After each acquisition, gesture segmentation was performed with a combination of manual inspection and an adaptive threshold to separate contractions from rest.

%used for the Unibo-INAIL dataset

\subsection{Multi-source data structure}\label{multisource}

The fundamental property of the Unibo-INAIL dataset is that it contains 224 different data sources, since the data were collected for all the combinations obtained from 7 seven subjects, 8 days and 4 arm postures. Each subject-day-arm posture combination contains all hand gestures (i.e.\ the classes), each one repeated 10 times. The dataset can thus be regarded as a collection of 224 autonomous sub-datasets, for which the pattern to learn in order to assign signals to hand gestures is subjected to inter-subject, inter-day and inter-posture variability.

In machine learning, data having this structure are defined multi-source data. For the Unibo-INAIL dataset, each combination identified by a 3-ple subject-day-arm posture combination is a source:
\begin{align}
\textrm{source} = (u, d, p) \quad \textrm{with} \begin{cases}
u &= 1, \cdots, 7\\
d &= 1, \cdots, 8\\
p &= 1, \cdots, 4\\\end{cases}
\end{align}
totalling 224 sources. Each source can be regarded as a smaller, complete dataset containing 10 repetitions of all the 6 gestures.\\
NOTE. In this work, the term \textit{multi-source data} is used according to the meaning it has in statistical learning \cite{Crammer2008}, where it refers to data-subsets having different but similar distributions. The intended meaning is not data coming from different types of sensors or modalities.

\begin{figure}[H]
  \centering
  \begin{subfigure}[b]{0.75\textwidth}
  \includegraphics[width=\textwidth]{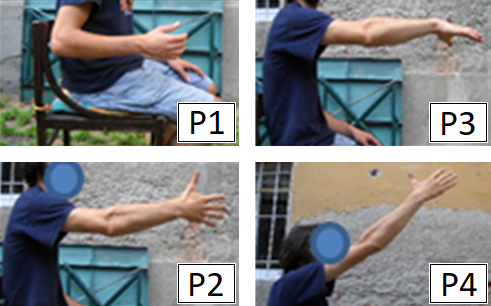}
  \end{subfigure}
  \caption{Arm postures of the Unibo-INAIL datase. P1: proximal; P2: distal; P3: distal with the palm oriented down; P4: distal with the arm lifted up by \SI{45}{\degree}. Image from \cite{Benatti2014}.}
  \label{img_dataset_postures}
\end{figure}

The subject-index $u$ refers to the 7 subjects involved. Each one underwent data collection over 8 days: this is the day-index $d$ of the data sources. Arm posture $p$ is the third index, and the four collected arm postures are:
\begin{itemize}\itemsep0em
\item[P1.] proximal: the only one with the arm not fully extended, and the most common in EMG-based hand gesture recognition literature;
\item[P2.] distal;
\item[P3.] distal with the palm oriented down: the different wrist orientation aims to introduce additional difference compared to P2 and P4;
\item[P4.] distal with the arm lifted up by \SI{45}{\degree}.
\end{itemize}
These arm postures are displayed in Figure \ref{img_dataset_postures}.

To constitute the classes of the dataset, five common hand gestures used in daily life were chosen: power grip, two-fingers pinch grip, three-fingers pinch grip, pointing index and open hand. Rest position, recorded when muscles were relaxed between two subsequent movement repetitions, was also included as a class, totalling 6 classes:
\begin{itemize}\itemsep0em
\item[$G_0$:] rest position: including this class means addressing also the task of gesture detection, in addition to gesture recognition;
\item[$G_1$:] power grip;
\item[$G_2$:] two-fingers pinch grip;
\item[$G_3$:] three-fingers pinch grip;
\item[$G_4$:] pointing index;
\item[$G_5$:] open hand.
\end{itemize}
The five hand gestures $G_1, \cdots, G_5$ are shown in Figure \ref{}, together with examples of 4-channel sEMG signal patterns of the gestures and the rest position $G_0$.\\
NOTE. It is of crucial importance, when handling the Unibo-INAIL dataset, not to mistake \textit{arm posture} and \textit{hand gesture}, since they are partitions acting at two different levels: the arm posture is the position of the arm in which all hand gestures (plus rest position) were executed. Each combination of subject, day and arm posture contains 10 repetitions of all the five hand gestures plus rest, and can thus be regarded as a complete sub-dataset containing all the 6 classes.

\subsection{User adaptation}\label{useradaptation}

User adaptation is the source of temporal variability which consists in a transient observed in the first days of many benchmarks datasets for sEMG-based hand gesture recognition \cite{Kaufmann2010, Asmuss2013, He2015}. The phenomenon consists in the fact that the inter-day differences in gesture execution decrease over time, due to the tendency of users to adapt to the repetitive exercise during the first days.

In literature, these inter-day differences are detected and measured by analysing how classification accuracy deteriorates when passing from intra-day validation to inter-day validation. With this method, classical machine leaning algorithms have already been able to highlight user adaptation also on the Unibo-INAIL dataset \cite{Milosevic2017}, which means that user the adaptation transient is not masked by the temporal variability caused by day-to-day sensor repositioning, which affects the signal on both the earlier and later days.

% rather than via descriptive analyses on the signals,

\begin{figure}[H]
  \centering
  \begin{subfigure}[b]{0.60\textwidth}
  \includegraphics[width=\textwidth]{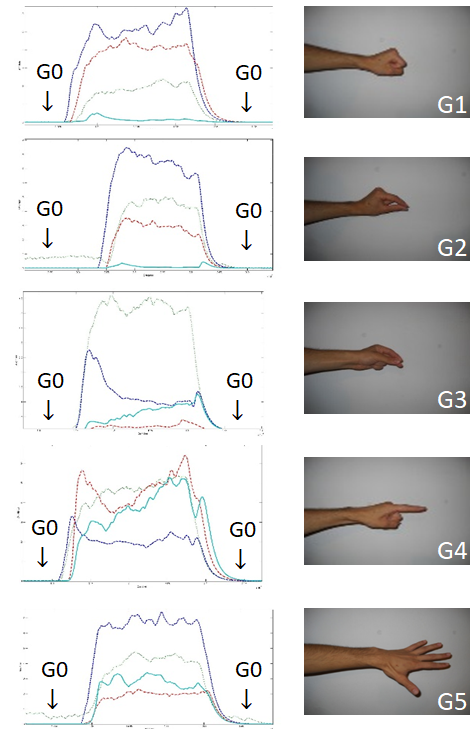}
  \end{subfigure}
  \caption{Amplified sEMG signal patterns of the hand gestures constituting the classes of the Unibo-INAIL dataset. $G_0$: rest position; $G_1$: power grip; $G_2$: two-fingers pinch grip; $G_3$: three-fingers pinch grip; $G_4$: pointing index; $G_5$: open hand. Image adapted from \cite{Benatti2014}.}
  \label{img_dataset_gestures}
\end{figure}

\newpage
\section{Pipeline and CNN architecture}\label{pipeline}

%...The machine learning pipeline implemented is typical of off-line sEMG signal recognition, and its steps are preprocessing, three-way data partition, training and validation of the CNNs ... variations of parameters... selection and test of the best one...

\subsection{Preprocessing: windowing}\label{preprocessing}

Data preprocessing consisted in an overlapping windowing scheme: the electromyogram signals were decomposed into segments of duration \SI{150}{\ms}, with a 75\% overlap between consecutive ones. Since the dataset was acquired with 4 electrodes and sampling rate \SI{500}{\Hz}, for approximately $\SI{15}{\min}/\textrm{session}$, this windowing produced a sample size of the order of $M \sim 25.000\textrm{ windows/session}$, each with dimensions $75\textrm{ samples}\times 4\textrm{ channels}$. Each window was given the label of the hand gesture of its central sample.

The choice of duration \SI{150}{\ms} and overlap 75\% was made with a preliminary analysis exposed in Subsection \ref{windowlength}, showing that duration \SI{150}{\ms} yields a higher classification accuracy than \SI{50}{\ms} and \SI{100}{\ms}.

Window duration and overlap are parameters strictly related to the engineering problem met in HMI development: real-time classification robustness, especially during transients. Longer windows allow to reduce the impact of transients, but the window length must be shorter than 300ms \cite{Hudgins1993} to satisfy real-time usage constraints. \SI{150}{\ms} is a favourable but realistic compromise.
% Moreover, in case that the relevant signal information occurs to be concentrated in shorter, central section, the CNN used is able to assign the window center the highest weights.
With regard to overlap, 75\% is a good compromise to produce an adequate sample size ($M \sim 25.000\textrm{ windows/session}$) without exceeding in redundancy.
% NOTE. It is worth to remark that augmentation via overlap is useful also when addressing real-time robustness. For the very reason that real-time usage constraints restrict large sample sizes only to off-line operations, training is always done off-line, and real-time operations are limited to calibration and inference.

\subsection{Three-way data partition}\label{datapartition}
Each of the 224 sessions (i.e.\ combinations of subject, day and arm posture) of the dataset was subjected to a three-way data partition.\\

\textbf{Random 10\% holdout.} First, a random 10\% of the signal windows were held out as test set. Reproduction of identical holdout at each execution was ensured by setting NumPy's pseudorandom seed to a fixed value. No stratification with respect to class was enforced. This test set was used in the very last step of the pipeline, to compute on new data:
\begin{itemize}\itemsep0em
\item the inter-posture test accuracy of the best postural training strategy;
\item the inter-day test accuracy of the best multi-day training strategy.
\end{itemize}
% The fraction 10\% was chosen because it is a suitable value for machine learning applications where sample size is below the big-data regime, (e.g. $M_\textrm{Big Data} \gtrsim 10^6$). The highest size in this work is $M_\textrm{five days} \sim 10^5$, reached when merging training data from day 1 to 5. 

\textbf{2-fold partition with gesture integrity.} After holdout, a 2-fold partition was applied on the remaining data (again, separately for each data session), to create two sets acting in turn as training set and validation set, in a 2-fold cross-validation scheme. The two folds were created starting from a 10-fold linear split, then putting the odd intervals into Fold 1 and the even intervals into Fold 2. This partition was chosen because it is the one yielding the best classification accuracy in \cite{Milosevic2017}, where it is called Training 50\%D (D meaning decimal). The motivation of this scheme is shuffling the 10 gesture repetitions while approximately preserving the integrity of each repetition.

\subsection{CNN architecture implemented}

The architecture of the CNN implemented is an adaptation of the CNN module of the attention-based hybrid CNN-RNN architecture proposed by Hu et al.\ in \cite{Hu2018} for sEMG-based gesture recognition on five public benchmark databases (not including the Unibo-INAIL dataset). More in detail, the hybrid architecture is a very large sequential model which stacks multiple parallel 2d-CNNs (identically trained), an LSTM, and an attention module. This model was chosen as a starting point due to its good performance (better than the state-of-the-art at publication) and to its modular structure, which is inspiring for exploring variations.

In this work, the 2d-CNN module from \cite{Hu2018} was taken and converted to a 1d-CNN. Conversion from 2d to 1d was needed to act on the time windows produced in the preprocessing step (Section \ref{preprocessing}), having dimensions $75\textrm{ samples}\times 4\textrm{ channels}$. For a CNN, this format corresponds to 1d images $75\times 1$ possessing 4 channels (or ``colors'').

\begin{figure}[H]
  \centering
  \begin{subfigure}[b]{1\textwidth}
  \includegraphics[width=\textwidth]{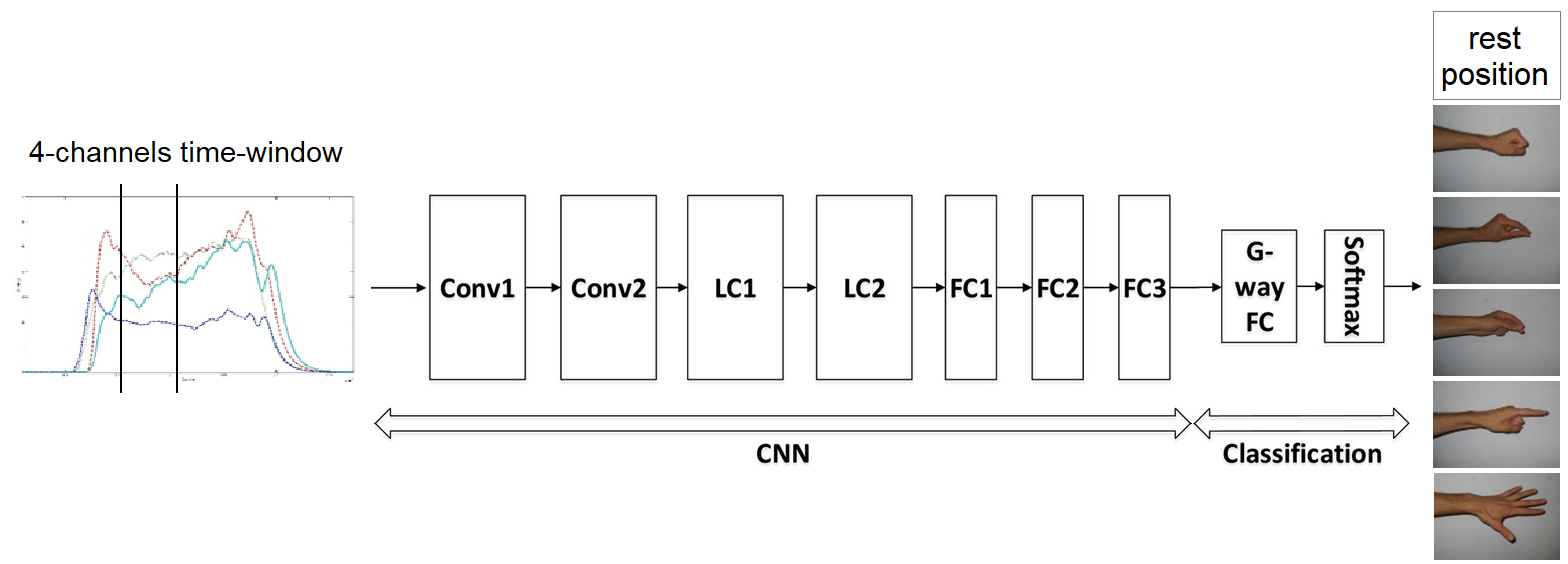}
  \end{subfigure}
  \caption{Diagram of the 1d-CNN implemented. Image adapted from \cite{Hu2018} and \cite{Benatti2017}.}\label{img_CNN}
\end{figure}

The resulting CNN architecture is shown in Figure \ref{img_CNN} and has 9 layers, listed in Table \ref{table_layers}. The first two layers are 1d-convolutional layers with 64 kernels of size 3. They are followed by two locally-connected layers with 64 kernels of size 1, employed to extract features of the sEMG signal that are temporally circumscribed (i.e.\ ``local'' in time). For all these layers, batch normalization is applied to reduce internal covariate shift. The fifth, sixth and seventh layers are all fully-connected layers with batch normalization. Moreover, dropout with probability $p = 0.5$ is applied to the first two fully-connected layers to provide regularization. The fully connected layers are followed by a 6-way fully-connected layer (6 being the number of classes, i.e.\ the five hand gestures plus the rest position) and a softmax classifier. Except for the latter, all layers have ReLU activation function.

\begin{table}[h]
\centering
\begin{tabular}{@{}ccc@{}}
\toprule
\textbf{Layer} & \textbf{Name} & \textbf{Details}                                                      \\ \midrule
1     & Conv1    & \begin{tabular}[c]{@{}c@{}}Convolutional 1d, 64 kernels, kernel size 3\end{tabular} \\
2     & Conv2    & \begin{tabular}[c]{@{}c@{}}Convolutional 1d, 64 kernels, kernel size 3\end{tabular} \\ \midrule
3     & LC1      & \begin{tabular}[c]{@{}c@{}}Locally connected, 64 kernels\end{tabular}               \\
4     & LC2      & \begin{tabular}[c]{@{}c@{}}Locally connected, 64 kernels\end{tabular}               \\ \midrule
5     & FC1      & \begin{tabular}[c]{@{}c@{}}Fully connected, 512 units\end{tabular}                \\
6     & FC2      & \begin{tabular}[c]{@{}c@{}}Fully connected, 512 units\end{tabular}                \\
7     & FC3      & \begin{tabular}[c]{@{}c@{}}Fully connected, 128 units\end{tabular}                \\ \midrule
8     & 6-Way FC & \begin{tabular}[c]{@{}c@{}}Fully connected, 6 units\end{tabular}                  \\
9     & SoftMax  & Softmax function                                                                    \\ \bottomrule
\end{tabular}
\caption{Layers of the 1d-CNN implemented.}
\label{table_layers}
\end{table}

\subsection{Training settings}
The optimal parameters regulating CNN training were found via preliminary analyses, observing the network's behaviour when trained with various settings. In particular, the optimal learning rate, number of epochs and scheduling were chosen through the preliminary analysis exposed in Subsection \ref{learningcurves}. The optimal settings, used for all the trainings in the pipeline, are the following:
\begin{itemize}\itemsep0em
\item random initialization of weights and biases: PyTorch default;
\item loss function: cross-entropy loss, implemented by the PyTorch command \texttt{criterion = torch.nn.CrossEntropyLoss()}, which stacks log-softmax (\texttt{nn.LogSoftmax()}) and negative log-likelihood loss (\texttt{nn.NLLLoss()});
\item optimization algorithm: Stochastic Gradient Descent (SGD) with the number of mini-batches kept fixed at $B = 50$ for all trainings (thus with mini-batch size varying proportionally to training set size); mini-batches are re-randomized at each epoch;
\item learning rate: $\texttt{lr} = 0.001$;
\item early stopping: 20 epochs;
\item scheduling: learning rate is divided by 10 after epoch 19;
\item regularization: $\text{L}_2$ regularization setting PyTorch parameter $\texttt{weight\_decay = 0.1}$, which corresponds to $\lambda_{\text{L}_2} = 0.05$.
\end{itemize}

\section{Training strategies}\label{trainingstrategies}

The very aim of this work is to assess the ability of the implemented CNN model to generalize to data coming from arm postures or days not seen in training. In particular, the focus is on evaluating whether training on more postures or more days benefits the CNN's generalization ability. This is done by implementing the same training strategies studied by Milosevic et al.\ in \cite{Milosevic2017}, i.e.\ single-session, two-posture, two-day, and five-day training. The positive results of Milosevic et al.\ (limited to classical machine learning classifiers applied to instantaneous 4-channel signal values) play the role of baseline.

\subsection{Single-session training strategy}

% For each session, i.e.\ for each combination of one subject, one day and one posture, a CNN was trained using only data coming from that session. The model's ability to generalize to different arm postures or days when trained on a single session was evaluated by validating each CNN on the other 3 arm postures (keeping fixed subject and day) and on the other 7 days (keeping fixed subject and arm posture).

The simplest training strategy implemented is the single-session training strategy: for each session, i.e.\ a combination of user, day, and arm posture, 2 CNNs are trained (one for each fold), using only data taken from that session. Classification accuracy is then evaluated via intra-session validation and inter-session validation:
\begin{itemize}\itemsep0em
\item in intra-session validation, the trained CNN is evaluated on the data of the fold not used for training;
\item in inter-session validation, the trained CNN is evaluated the data of a different session, to measure the CNN's ability to generalize; for the single-session training strategy, the inter-session validations computed are the inter-posture validation (all combinations), and inter-day validations on days D2 to D8 after training on day D1.
\end{itemize}

\subsection{Two-posture training strategy}

% To see how training on two postures affects the model's ability to generalize on different postures, CNNs were trained using combinations of two postures: $\textrm{P}1 + \textrm{P}2$, $\textrm{P}1 + \textrm{P}3$, and $\textrm{P}1 + \textrm{P}3$ (for each subject, for each day). Each model was then validated on the two postures not used for training (keeping fixed subject and day). The reason why P1 was used for all trainings, is that P1 is the only posture with the arm not extended (see Section \ref{multisource}), and therefore is suitable to maximize variability within the training set.

The training strategy developed to address postural variability is two-posture training: for each subject, day, and arm posture pair, 2 CNNs are trained (one for each fold), using only data taken from those two sessions (i.e.\ that subject, that day and those two postures). The posture pair considered are P$1+$P2, P$1+$P3, and P$1+$P4, where the choice of always including P1 is motivated by the fact that P1 is the only posture with the arm not fully extended, thus the most dissimilar from the other ones. Classification accuracy is then measured via intra-postures validation and inter-posture validation:
\begin{itemize}\itemsep0em
\item in intra-postures validation, the trained CNN is evaluated on the fold not used for training (the plural intra-posture\textit{s} is because each fold now contains data from two arm postures);
\item in inter-posture validation, the model is evaluated on the data of a different posture, to measure the model's ability to generalize between postures.
\end{itemize}

\subsection{Multi-day training strategies}

The training strategies proposed to address temporal variability is multi-day training: for each subject, day combination, and arm posture, 2 CNNs are trained (one for each fold), using only data from that subject, those days and that posture. In particular, multi-day trainings are two-day trainings, which use D$1+$D2, D$1+$D5, and D$4+$D5, and five-day training, which use D1 to D5. These combinations were chosen in order to repeat the setup which produced the baseline results. Classification accuracy is then measured via intra-day validation and inter-days validation:
\begin{itemize}\itemsep0em
\item in intra-days validation, the trained CNN is evaluated on the fold not used for training (the plural intra-posture\textit{s} is because each fold now contains data from two or five days);
\item in inter-day validation, the model is evaluated on the data of days D6 to D8 (absent in all multi-day training combinations), to measure the model's ability to generalize to different days.
\end{itemize}

%\ subsection{Two-day training strategy}
%To see how training on two days affects the model's ability to generalize on different days, CNNs were trained using combinations of two days: $\textrm{D}1 + \textrm{D}2$, $\textrm{D}1 + \textrm{D}5$, and $\textrm{D}4 + \textrm{D}5$ (for each subject, for each posture). Each model was then validated on days 6, 7, and 8 (keeping fixed subject and day). These particular day pairs were chosen to see if training on earlier or recent days yields some difference.

%\ subsection{Five-day training}
%A procedure analogous to two-day training was followed, using days 1 do 5 and testing on days 6, 7, and 8, in order to see how training on all sessions (both early and recent) prior to day 6 affects the generalization performance on new days.

% % % % % % % % % % % % % % % % % % % % % % % % % % % % % % % % % % % % % % % % % % % % % % % % % % % % % % % % % % % % % % % % % % % % % % % % % % % % % % % % % % % % % % % % % % % % % % % % % % % % % % % % % % % % % % % % % % % % % % % % % % % % % % % % % % % % % % % % % % % % % % % % % % % % % % % % % % % % % % % % % % % % % % % % % % % % % % % % %

\newpage
\chapter{Implementation}\label{implementation}

The pipeline and the CNN described in the previous chapter were implemented in Python scripts whose heart, dealing with CNN definition, training and evaluation, relies on the open source library PyTorch.

This chapter is structured as follows:
\begin{itemize}\itemsep0em
\item[$\circ$] Section \ref{scripts} illustrates in general the scripts developed for the different training strategies;
\item[$\circ$] Section \ref{PyTorch} explains more in detail the main PyTorch packages and how they were used.
\end{itemize}

\section{Scripts developed}\label{scripts}
Although the Unibo-INAIL dataset is publicly available with a series of Matlab scripts to help analyses, these were not exploited. Only one was partially used: the script \texttt{script\_training\_single\_sessions.m}, devoted to training all the classical machine learning tested in \cite{Milosevic2017} with single-session data, corresponding to the first training strategy (i.e.\ data from one subject, one day and one arm posture; details in Section \ref{trainingstrategies}). This script was translated into Python and the following advances were implemented:
\begin{itemize}\itemsep0em
\item data partition (detailed in Section \ref{pipeline}) was rewritten from scratch as the function \texttt{load\_preprocess\_and\_split(\_)}, to make it more compact, to add holdout and to make the $K$-fold partition easily costumizable by simply setting a parameter \texttt{K};
\item windowing preprocessing (see Section \ref{preprocessing}) was added as the function \texttt{windowing(\_)} (called inside the previous one, immediately before doing data partition), since the previous implementations only addressed the classifications of 4-channel instantaneous signal values;
\item in the core block, devoted to instantiation, training and validation, the four classical algorithms were replaced by the CNN, implemented with the library PyTorch.
\end{itemize}
The structure of \texttt{for} loops cycling on subjects, days and arm postures was approximately preserved.

This revised script was in turn used as a template to implement the two-posture, two-day and five-day training strategies (see Section \ref{trainingstrategies}), each one in a separate script, totalling four scripts:
\begin{itemize}\itemsep0em
\item \texttt{CNN\_on\_UniboINAIL.py}, for single-session trainings;
\item \texttt{biposture.py}, for two-posture trainings;
\item \texttt{biday.py}, for two-day trainings;
\item \texttt{fiveday.py}, for five-day trainings.
\end{itemize}
Execution times on a single GPU were approximately \SI{8}{\hour} for \texttt{CNN\_on\_UniboINAIL.py} and \texttt{biposture.py}, and approximately \SI{5}{\hour} for \texttt{biday.py} and \texttt{fiveday.py}.

\section{Usage of PyTorch platform}\label{PyTorch}
The CNN instantiation, training and validation were implemented using the Python library PyTorch, an open source deep learning platform for Python, based on Torch, that is widely used in deep learning applications. This work has taken advantage of both PyTorch's high-level features: tensor computation on variables of class \texttt{torch.Tensor} (partially analogous to NumPy arrays, but more powerful), and the possibility of GPU computation.

In this master thesis, the CNNs were implemented in compliance with the standard structure of PyTorch scripts for CNNs, and the main PyTorch packages exploited are \texttt{torch.nn}, for neural networks instantiation and usage, \texttt{torch.autograd}, for automatic differentiation, \texttt{torch.optim}, for optimization, and \texttt{torch.cuda} for GPU computation. The following subsections explain these packages and the way they were used.

\subsection{\texttt{torch.nn} package}
\texttt{torch.nn} is the package that helps defining the complex neural networks for which operations on \texttt{torch.Tensor}s with raw \texttt{autograd} alone are too low-level. In particular, \texttt{torch.nn.Module} is the base class for all neural network modules, and user-defined models must subclass this class. If desired, modules (submodules) can be nested inside other Modules, creating a tree structure.

While reporting the code written to define the CNN architecture would be excessive, it is worth to show how instantiating the defined architecture is straightforward. It is done via two commands:
\begin{flushleft}
%\texttt{net = nn.DataParallel(Net().to(device))}\\
\texttt{net = Net()}\\
\texttt{criterion = torch.nn.CrossEntropyLoss()}
\end{flushleft}
The first line instantiates the CNN \texttt{net}. The second line instantiates the loss function, termed criterion in PyTorch terminology. The class \texttt{torch.nn.CrossEntropy} implements a cross-entropy loss function which stacks a log-softmax operation (\texttt{torch.nn.LogSoftmax()}) and negative log-likelihood loss (\texttt{torch.nn.NLLLoss()}).

\subsection{\texttt{torch.autograd} package and \texttt{torch.Tensor} class}
\texttt{torch.autograd} is the package that implements automatic differentiation: gradients with respect to the parameters are automatically calculated directly at the forward pass (by recording the executed operations into computational graphs and re-executing them backward), thus reducing both development and execution time. The classes and functions provided implement automatic differentiation of arbitrary scalar-valued functions, with the only requirement that the variables with respect to which gradients are computed be \texttt{torch.Tensor} objects.

\texttt{torch.Tensor} is the base class of the \texttt{autograd} package. \texttt{torch.Tensor}s are multi-dimensional arrays (similar to NumPy array), which support automatic differentiation. Every \texttt{torch.Tensor} variable possesses a \texttt{requires\_grad} flag, which allows to enable or disable gradient computation with respect to that variable. So, typically, data are \texttt{torch.Tensor}s with \texttt{requires\_grad=False}, and model parameters are \texttt{torch.Tensor}s with \texttt{requires\_grad=True}. Disabling gradients is also useful for freezing parts of a model, e.g.\ when fine-tuning other model parts.

The operations performed on tensors having \texttt{requires\_grad} set to \texttt{True} are tracked. After finishing computations, gradients can be computed automatically by calling \texttt{backward()} on the result, i.e.\ the variable whose value was computed with the function to differentiate. The gradient with respect each \texttt{torch.Tensor} is then accumulated in its \texttt{.grad} attribute\footnote{Formally, differentiation computes gradients \textit{of} functions \textit{with respect to} arguments (here, \texttt{torch.Tensor} parameters). Sometimes, in machine learning and deep learning, the abuse of language is made of speaking of derivatives \textit{of} the parameters. The justification is that differentiation almost always acts on the training objective function. However, in \texttt{autograd}, speaking of gradients \textit{of} the tensors is correct in the sense that gradient values are stored in each tensor's \texttt{.grad} attribute.}.

Formally, \texttt{autograd} is a reverse automatic differentiation system. As manipulations on data are executed, \texttt{autograd} produces a graph recording the operations that originate the result. This yields a directed acyclic graph having input tensors as leaves and output tensors as roots. Gradients are computed automatically by chain rule, by tracing the computational graph from roots to leaves (thus executing back-propagation).

Inside \texttt{autograd}, computational graphs are represented as graphs of \texttt{torch.Function} objects. During the forward pass, \texttt{autograd} simultaneously creates the graph representing the function that computes the gradient. Then, in the backward pass the graph is evaluated to compute the gradients. A valuable property is that the graph is re-built from scratch at every iteration, which allows to use arbitrary Python control flow statements, that can originate different graph structures at every iteration. This is the define-by-run framework, in which there is no need to encode all possible paths before launching the training, also described with the sentence ``What you run is what you differentiate''.%\\
%NOTE ON TERMINOLOGY. Formally, what the \texttt{autograd} engine computes are Jacobian matrices in general. The reason why the outputs are commonly referred to as \textit{gradients} is that in machine leaning and deep learning the function to differentiate is almost always only the training objective function, which is a scalar function.

The \texttt{torch.autograd} mechanics were exploited to compute the gradient of the training loss, computed as a cross-entropy. After initializing the model and the cross-entropy objective function by \texttt{net = Net()} and
\texttt{criterion = torch.nn.CrossEntropyLoss()} (as explained in the previous subsection), the values of the loss and loss gradient were computed at each iteration as follows:
\begin{flushleft}
%\texttt{loss = criterion(net(XTrain[batch\_idxs, :, :].to(device)), YTrain[batch\_idxs].to(device))}\\
\texttt{loss = criterion(net(XTrain[batch\_idxs]), YTrain[batch\_idxs])}\\
\texttt{loss.backward()}
\end{flushleft}

\subsection{\texttt{torch.optim} package}
The package \texttt{torch.optim} implements the most common iterative optimization algorithms. It is used instantiating and handling an \texttt{Optimizer} object (or more than one), which keeps track of the current state of optimization and, when its \texttt{step()} method is called, updates the parameters based on the gradients previously computed.

The optimization algorithm used in this work is Stochastic Gradient Descent (SGD) with mini-batches (re-randomized at each epoch), implemented via the command
\begin{flushleft}
\texttt{optimizer = torch.optim.SGD(net.parameters(), lr=0.001, weight\_decay=0.1)}
\end{flushleft}
where \texttt{SGD(\_)} is the \texttt{torch.optim} function which instantiates SGD optimizers (without requiring to declare mini-batch size), \texttt{net.parameters()} are intuitively the parameters of the CNN model \texttt{net}, \texttt{lr} is the learning rate, and \texttt{weight\_decay} corresponds to $2\cdot\lambda_{\text{L}_2}$ determining the amount of $\text{L}_2$ regularization.

For each mini-batch, optimization steps are executed by calling the \texttt{step()} method of the optimizer, immediately after computing the gradients on the mini-batch via \texttt{backward()}:
\begin{flushleft}
%\texttt{loss = criterion(net(XTrain[batch\_idxs, :, :].to(device)), YTrain[batch\_idxs].to(device))}\\
\texttt{loss = criterion(net(XTrain), YTrain)}\\
\texttt{loss.backward()}\\
\texttt{optimizer.step()}
\end{flushleft}
This performs one update of the parameters, based on the gradients.

To implement scheduling, \texttt{torch.optim.lr\_scheduler} was used, which provides methods for non-dynamic scheduling, i.e.\ learning rate adjustment based solely on epoch number, without adaptive validation measurements. The scheduling chosen after preliminary analysis (results in Subsection \ref{learningcurves}), i.e.\ division of the learning rate by 10 at epoch 19, was implemented via the command:
\begin{flushleft}
\texttt{scheduler = torch.optim.lr\_scheduler.StepLR(optimizer,}\\
\texttt{\qquad\qquad\qquad\qquad\qquad\qquad\qquad\qquad\qquad\qquad\qquad step\_size=19, gamma=0.1)}
\end{flushleft}
where \texttt{optimizer} is the optimizer (in this case, a \texttt{torch.optim.SGD} object) whose learning rate is being scheduled, and \texttt{step\_size} and \texttt{gamma} are the period and the multiplicative factor of the learning rate decay, respectively. Then, scheduling is applied simply by calling \texttt{scheduler.step()} at each epoch.

\subsection{\texttt{torch.CUDA} package}

In addition to supporting automatic differentiation, a further advantage of \texttt{torch.Tensor}s over NumPy arrays is that they can be cast to a GPU to improve computational performance.

This can be made using the package \texttt{torch.cuda}. It keeps track of the currently selected GPU, and all allocated CUDA tensors are created  by default on that device. A \texttt{torch.cuda.device} context manager allows to change the selected device. However, once a \texttt{torch.Tensor} is allocated, operations can be performed on it irrespective of the selected device, and the results are automatically placed on the same device as the \texttt{torch.Tensor}. A \texttt{torch.Tensor}'s device can be accessed via the \texttt{Tensor.device} property.

The GPU computation is enabled by adding
\begin{flushleft}
\texttt{device = torch.device('cuda:0' if torch.cuda.is\_available() else 'cpu')}
\end{flushleft}
at the beginning of the script, to set \texttt{device} to \texttt{'cuda:0'} if a GPU device is available. A \texttt{torch.device} is an object representing the device on which a \texttt{torch.Tensor} is or will be allocated, and is constructed so as to contain the device type (\texttt{'cpu'} or \texttt{'cuda'}) and an optional device ordinal for the device type.

The changes to cast model and data (and thus computations) to the GPU are few and straightforward, and are made via the command \texttt{.to(device)}. The CNN instantiation \texttt{net = Net()} becomes
\begin{flushleft}
\texttt{net = Net().to(device)}
\end{flushleft}
and the evaluation and loss computation \texttt{loss = criterion(net(XTrain), YTrain)} becomes
\begin{flushleft}
\texttt{loss = criterion(net(XTrain[batch\_idxs].to(device)),}\\
\texttt{\qquad\qquad\qquad\qquad\phantom{ }YTrain[batch\_idxs].to(device))}
\end{flushleft}

% % % % % % % % % % % % % % % % % % % % % % % % % % % % % % % % % % % % % % % % % % % % % % % % % % % % % % % % % % % % % % % % % % % % % % % % % % % % % % % % % % % % % % % % % % % % % % % % % % % % % % % % % % % % % % % % % % % % % % % % % % % % % % % % % % % % % % % % % % % % % % % % % % % % % % % % % % % % % % % % % % % % % % % % % % % % % % % % %

\chapter{Results}\label{results}

\section{Accuracy distributions and reported accuracies}

Due to the highly multi-source nature of the Unibo-INAIL dataset, the CNNs trained with each training strategy (single-session, two-postures, two- and five-day) return a distribution of classification accuracies, over the 7 subjects, 8 days and 4 arm postures. This allows to investigate the variability of performance over the dataset. On this basis, in this Chapter all accuracies are reported with:
\begin{itemize}\itemsep0em
\item average accuracy $\mu$ (also called mean for simplicity), computed over the sessions of interest for each case;
\item accuracy standard deviation $\sigma$, which quantifies the performance variability over the data sources;
\item standard error on the mean accuracy $\text{SE} = \sigma/\sqrt{N}$, used to estimate the uncertainty on the average accuracy.
\end{itemize}
In particular, the SE does not describe the broadness of the accuracy distribution, but is an estimate of the fluctuations affecting the average accuracy. The 224 sessions (times 2 folds) of the Unibo-INAIL dataset allow to divide by a large $\sqrt{N}$ (varying according to the training strategy studied). Thus the 224 sessions are not only a computational burden, but also allow to report average accuracies with a SE of the order of $\sim 0.1\%$.

In this chapter, the reported accuracies are compared with the baseline achieved by the best classical machine learning classifier reported in \cite{Milosevic2017}. However, for the baseline accuracies the values of $\sigma$ or SE are not reported, not allowing for a complete statistical comparison.

\section{Preliminary analyses}

\subsection{Optimal length of time windows}\label{windowlength}

The first preliminary analysis was made to optimize the window length to be used in preprocessing, consisting in an overlapping windowing scheme as detailed in Subsection \ref{preprocessing}. Since window length must be shorter than 300ms to satisfy real-time usage constraints, the values explored were \SI{50}{\ms}, \SI{100}{\ms} and \SI{150}{\ms}.

In this analysis, the performance of interest is the accuracy obtained on 2-fold cross-validation on single-session data (i.e. single user, single day, single arm posture), after training on data from the same session (i.e.\ using the two folds alternately for training and intra-session validation). The results are reported in Table \ref{table_windows}. With a $(94.4 \pm 0.2)\%$ accuracy, duration \SI{150}{\ms} proved to be the best one and was adopted the preprocessing.

Since the dataset was acquired with 4 electrodes and sampling rate \SI{500}{\Hz}, for approximately $\SI{15}{\min}/\textrm{session}$, this windowing produced a sample size of the order of $M \sim 25.000\textrm{ windows/session}$, each with dimensions $75\textrm{ samples}\times 4\textrm{ channels}$. Moreover, with regard to overlap, 75\% was chosen as a good compromise to produce an adequate sample size ($M \sim 25.000\textrm{ windows/session}$) without exceeding in redundancy.

\begin{table}[H]
\centering
\begin{tabular}{@{}ccc@{}}
\toprule
\multirow{2}{*}{\textbf{Window length}} & \multicolumn{2}{c}{\textbf{Intra-session val. accuracy}} \\
                               & $\mu\pm\text{SE}$      & $\sigma$     \\ \midrule
\phantom{1}\SI{50}{\ms}                & $(93.6\pm 0.2)\%$              & $3.7\%$      \\
\SI{100}{\ms}                          & $(94.0\pm 0.2)\%$              & $3.6\%$      \\
\SI{150}{\ms}                          & $(94.4\pm 0.2)\%$               & $3.5\%$      \\ \bottomrule
\end{tabular}
  \caption{Intra-session validation accuracies yielded by the time-windows durations explored.}
  \label{table_windows}
\end{table}

\subsection{Learning curves}\label{learningcurves}

The second preliminary analysis was performed to optimize the learning rate and the scheduling thereof. This exploration was carried on with the same single-session scheme used to optimize the window length described in Subsection \ref{windowlength}.

\begin{figure}[H]
  \centering
  \begin{subfigure}[b]{0.6\textwidth}
  \includegraphics[width=\textwidth]{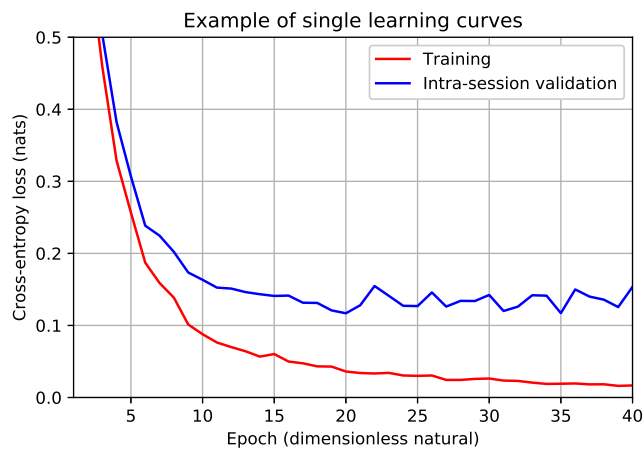}
  \end{subfigure}
  \caption{Single learning curves, obtained for subject 1, day 1, arm posture 1, training on fold 1 with learning rate $\texttt{lr} = 0.001$, validation on fold 2.}
  \label{img_curves_1}
\end{figure}

The optimal learning rate was searched by looking at single intra-session validation learning curves, like the one shown in Figure \ref{img_curves_1} (obtained for subject 1, day 1, arm posture 1, training on fold 1 and validation on fold 2). Using SGD with the training set split into 50 mini-batches (re-randomized at each epoch), the optimal learning rate was found to be $\texttt{lr} = 0.001$, which yielded the best compromise between a reasonably fast convergence and a sufficiently low final loss. This optimal value was adopted for all the trainings.

\begin{figure}[H]
  \centering
  \begin{subfigure}[b]{1\textwidth}
  \includegraphics[width=1\textwidth]{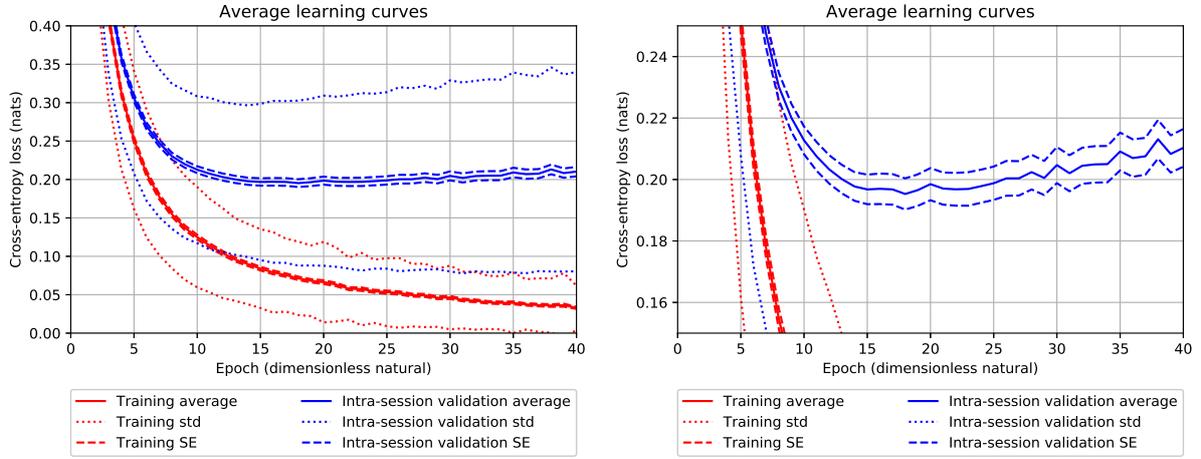}
  \end{subfigure}
  \caption{Average learning curves obtained with learning rate $\texttt{lr} = 0.001$, no scheduling. Averages, $\sigma$'s and SEs taken on all subjects, all days, all arm postures, both folds. The left and right panels show the same learning curves, visualized with $y$-axis range $[0, 0.4]$, and $[0.15, 0.25]$, respectively.}
  \label{img_curves_2}
\end{figure}

\begin{figure}[H]
  \centering
  \begin{subfigure}[b]{1\textwidth}
  \includegraphics[width=1\textwidth]{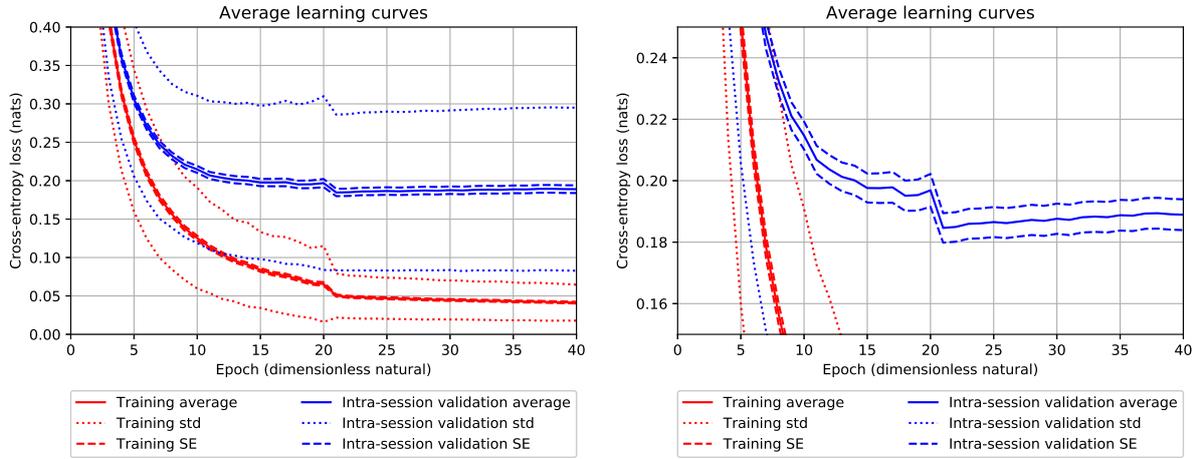}
  \end{subfigure}
  \caption{Average learning curves obtained with learning rate $\texttt{lr} = 0.001$, divided by 10 at epoch 20. Averages, $\sigma$'s and SEs taken on all subjects, all days, all arm postures, both folds. The left and right panels show the same learning curves, visualized with $y$-axis range $[0, 0.4]$, and $[0.15, 0.25]$, respectively.}
  \label{img_curves_3}
\end{figure}

Average learning curves were used to diagnose overfitting, as shown in Figure \ref{img_curves_2}: the validation loss reaches a minimum between epochs 15 and 20, than increases again, indicating overfitting. A simple scheduling tactic, consisting in dividing the learning rate by 10 at epoch 20, allowed to improve the minimum of the validation loss, as shown in Figure \ref{img_curves_3}. Since this scheduling does not remove the overfitting trend, but only makes it slower, the final choice was to exploit the sudden minimum: the final scheduling chosen consists in dividing the learning rate by 10 at epoch 19, then applying early stopping at epoch 20. This scheduling was used in all the trainings.\\
NOTE. After this preliminary analysis, validation cross-entropy losses were abandoned in favour of classification accuracies. This was motivated by the ease of interpretation and informal comparison with baseline accuracies.

\newpage
\section{Single-session training strategy}

The basic training strategy explored is the single-session training strategy: for each session, i.e.\ a combination of user, day, and arm posture, 2 CNNs are trained (one for each fold), using only data from that session. Performance is then measured via intra-session validation and inter-session validation:
\begin{itemize}\itemsep0em
\item in intra-session validation, the trained CNN is evaluated on the fold not used for training;
\item in inter-session validation, the model is evaluated on the data of a different session, to assess the model's ability to generalize; for the single-session training strategy, the inter-session validations computed are the inter-posture validation (all combinations), and inter-day validations on days D2 to D8 after training on day D1.
\end{itemize}

With regard to intra-session validation, the overall accuracy is $(\mu \pm \text{SE}) = (94.5 \pm 0.2)\%$, with $\sigma = 3.5\%$. This performance is very similar to the baseline value of 94\% achieved with a RBF-SVM, which is not consistent within the SE, but a complete statistical comparison is not possible because the baseline $\sigma$ is not available.

\begin{figure}[H]
  \centering
  \begin{subfigure}[b]{0.8\textwidth}
  \includegraphics[width=1\textwidth]{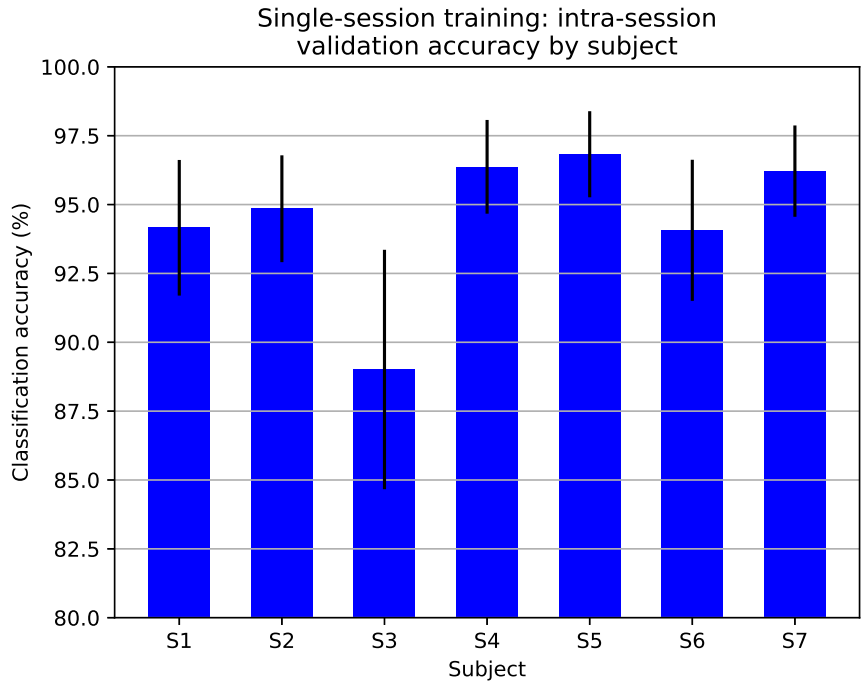}
  \end{subfigure}
  \caption{Single-session training: intra-session validation accuracies by subject. Error bars stand for $\pm\sigma$}
  \label{img_ssacc_U}
\end{figure}

\begin{figure}[H]
  \centering
  \begin{subfigure}[b]{0.8\textwidth}
  \includegraphics[width=1\textwidth]{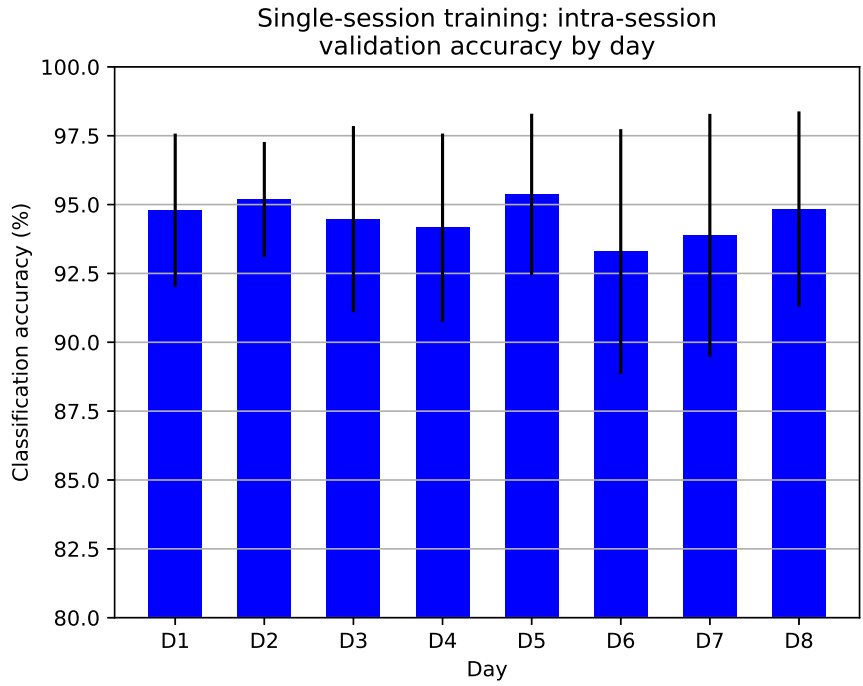}
  \end{subfigure}
  \caption{Single-session training: intra-session validation accuracies by day. Error bars stand for $\pm\sigma$}
  \label{img_ssacc_D}
\end{figure}

\begin{figure}[H]
  \centering
  \begin{subfigure}[b]{0.8\textwidth}
  \includegraphics[width=1\textwidth]{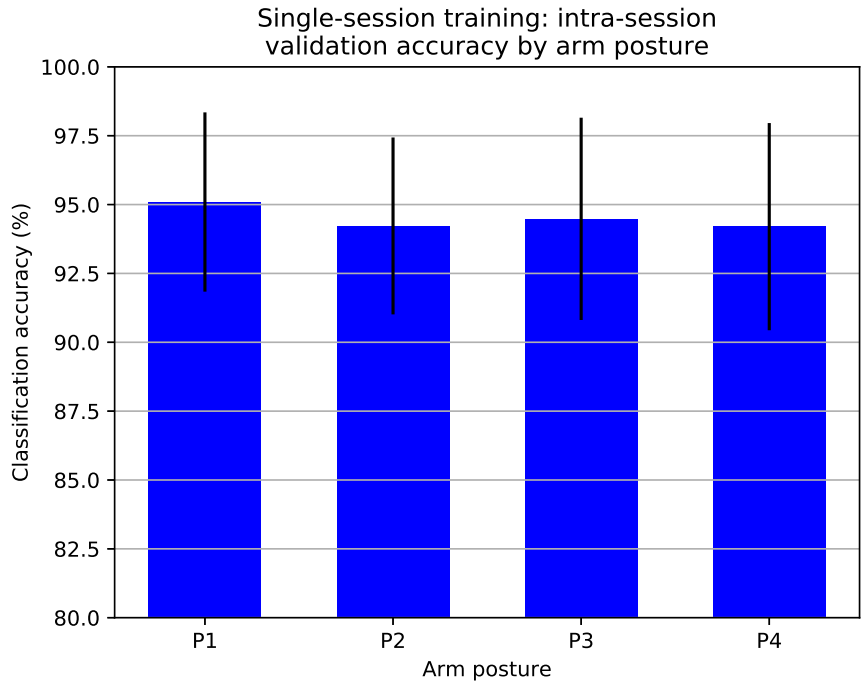}
  \end{subfigure}
  \caption{Single-session training: intra-session validation accuracies by arm posture. Error bars stand for $\pm\sigma$}
  \label{img_ssacc_P}
\end{figure}

Looking at accuracy distributions computed by subject, by day and by arm posture, some interesting findings emerge. Accuracy distributions by subject are reported in Figure \ref{img_ssacc_U} and in Table \ref{table_ssacc_U}; accuracy distributions by day are reported in Figure \ref{img_ssacc_D} and in Table \ref{table_ssacc_D}; and accuracy distributions by arm posture are reported in Figure \ref{img_ssacc_P} and in Table \ref{table_ssacc_P}.

The accuracy distributions (visualized as $\mu\pm\sigma$) of the 8 days and of the 4 arm postures always overlap within the standard deviations. The situation is different for the accuracy distributions for the 8 subjects: the means present larger variations, and Subject 3 has a mean so lower than the others that its accuracy distribution is not consistent within the standard deviation with the distributions of Subjects 4, 5, and 7. Subject 3's accuracy distribution is also the one presenting the highest standard deviation. Operatively, this indicates that, on average, within Subject 3's sessions hand gesture recognition is a harder task. It is worth to remark that the inter-session setup does not authorize to attribute lower accuracy to larger inter-day or inter-postural variability.

% Interpretation of sigmas: users < days < postures.

\begin{figure}[H]
  \centering
  \begin{subfigure}[b]{1\textwidth}
  \includegraphics[width=1\textwidth]{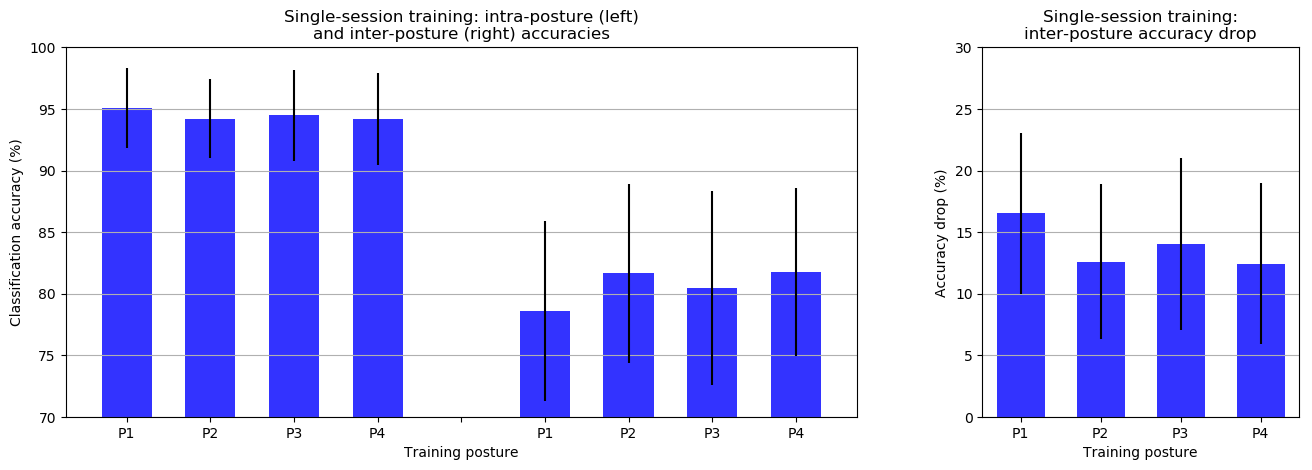}
  \end{subfigure}
  \caption{Single-session training: intra-posture and inter-posture validation accuracies, and accuracy drops, by training posture. Error bars stand for $\pm\sigma$}
  \label{img_ssPdrop}
\end{figure}

With regard to inter-session validation, Figure \ref{img_ssPdrop} and Table \ref{img_ssPdrop} show the inter-posture validation accuracies and compare them with the intra-posture case; all performances are reported by training posture. The overall inter-posture validation accuracy is 80.6\%, corresponding to a 13.9\% accuracy drop compared to the intra-posture (i.e.\ intra-session) scenario. This accuracy drop quantifies the amount of overfitting the single-session training produces with respect to the task of generalizing to different postures. The overall inter-posture accuracy is similar to the baseline value of 79\%, but, again, a statistical comparison is not possible because the baseline $\sigma$ or SE are not available.

\newpage

\begin{table}[H]
\centering
\begin{tabular}{@{}ccc@{}}
\toprule
\multirow{2}{*}{\textbf{Subject}} & \multicolumn{2}{c}{\textbf{Intra-session val. accuracy}} \\
                         & $\mu \pm \text{SE}$                  & $\sigma$          \\ \midrule
1                        & $(94.2 \pm 0.3)$\%            & 2.5\%           \\
2                        & $(94.8 \pm 0.2)$\%           & 1.9\%           \\
3                        & $(89.0 \pm 0.5)$\%            & 4.3\%           \\
4                        & $(96.4 \pm 0.2)$\%            & 1.7\%           \\
5                        & $(96.8 \pm 0.2)$\%            & 1.6\%           \\
6                        & $(94.1 \pm 0.3)$\%            & 2.6\%           \\
7                        & $(96.2 \pm 0.2)$\%            & 1.7\%          \\ \bottomrule
\end{tabular}
\caption{Single-session training: intra-session validation accuracies by subject.}
\label{table_ssacc_U}
\end{table}

\newpage

\begin{table}[H]
\centering
\begin{tabular}{@{}ccc@{}}
\toprule
\multirow{2}{*}{\textbf{Day}} & \multicolumn{2}{c}{\textbf{Intra-session val. accuracy}} \\
                     & $\mu \pm \text{SE}$                 & $\sigma$           \\ \midrule
1                    & $(94.8 \pm 0.4)$\%           & 2.8\%            \\
2                    & $(95.2 \pm 0.3)$\%           & 2.1\%            \\
3                    & $(94.5 \pm 0.4)$\%           & 3.4\%            \\
4                    & $(94.2 \pm 0.4)$\%           & 3.4\%            \\
5                    & $(95.4 \pm 0.4)$\%           & 2.9\%            \\
6                    & $(93.3 \pm 0.6)$\%           & 4.4\%            \\
7                    & $(93.9 \pm 0.6)$\%           & 4.4\%            \\
8                    & $(94.8 \pm 0.4)$\%           & 3.5\%            \\ \bottomrule
\end{tabular}
\caption{Single-session training: intra-session validation accuracies by day.}
\label{table_ssacc_D}
\end{table}

\newpage

\begin{table}[H]
\centering
\begin{tabular}{@{}ccc@{}}
\toprule
\multirow{2}{*}{\textbf{Arm posture}} & \multicolumn{2}{c}{\textbf{Intra-session val. accuracy}} \\
                             & $\mu \pm \text{SE}$                 & $\sigma$           \\ \midrule
1                            & $(95.1 \pm 0.3)$\%           & 3.3\%            \\
2                            & $(94.2 \pm 0.3)$\%           & 3.2\%            \\
3                            & $(94.5 \pm 0.3)$\%           & 3.7\%            \\
4                            & $(94.2 \pm 0.4)$\%           & 3.8\%           \\ \toprule
\end{tabular}
\caption{Single-session training: intra-session validation accuracies by arm posture.}
\label{table_ssacc_P}
\end{table}

\newpage

\begin{table}[H]
\centering
\small
\begin{tabular}{@{}ccccccc@{}}
\toprule
\multirow{2}{*}{\textbf{Train posture}} & \multicolumn{2}{c}{\textbf{Intra-posture accuracy}} & \multicolumn{2}{c}{\textbf{Inter-posture accuracy}} & \multicolumn{2}{c}{\textbf{Accuracy drop}} \\
                                  & $\mu \pm \text{SE}$                & $\sigma$         & $\mu \pm \text{SE}$                & $\sigma$         & $\mu \pm \text{SE}$           & $\sigma$     \\ \midrule
1                                 & $(95.1 \pm 0.3)\%$          & 3.3\%          & $(78.6 \pm 0.4)\%$          & 7.3\%          & $(16.5 \pm 0.4)\%$     & 6.5\%      \\
2                                 & $(94.2 \pm 0.3)\%$          & 3.2\%          & $(81.6 \pm 0.4)\%$          & 7.3\%          & $(12.6 \pm 0.3)\%$     & 6.3\%      \\
3                                 & $(94.5 \pm 0.3)\%$          & 3.7\%          & $(80.5 \pm 0.5)\%$          & 8.9\%          & $(14.0 \pm 0.4)\%$     & 7.0\%      \\
4                                 & $(94.2 \pm 0.4)\%$          & 3.8\%          & $(81.8 \pm 0.4)\%$          & 6.8\%          & $(12.4 \pm 0.4)\%$     & 6.6\%      \\ \bottomrule
\end{tabular}
  \caption{Single-session training: intra-posture and inter-posture validation accuracies.}
  \label{table_ssPdrop}
\end{table}

\newpage
\section{Two-posture training strategy}

\begin{figure}[H]
  \centering
  \begin{subfigure}[b]{1\textwidth}
  \includegraphics[width=1\textwidth]{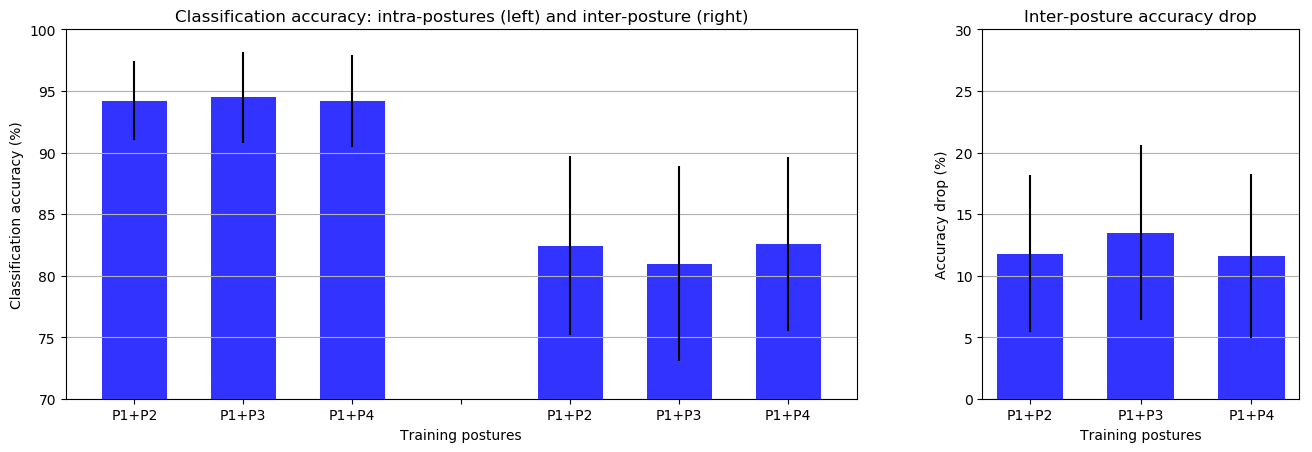}
  \end{subfigure}
  \caption{Two-posture training: intra-postures and inter-posture validation accuracies, and accuracy drops, by training posture pair. Error bars stand for $\pm\sigma$}
  \label{img_bpPdrop}
\end{figure}

The training strategy implemented to address postural variability is two-posture training: for each subject, day, and arm posture pair, 2 CNNs are trained (one for each fold), using only data from those two sessions (i.e.\ that subject, that day and those two postures). Posture pair considered are P$1+$P2, P$1+$P3, and P$1+$P4, where the choice of always including P1 is motivated by the fact that P1 is the only posture with the arm not fully extended, thus the most dissimilar from the other ones. Performance is then measured via intra-postures validation and inter-posture validation:
\begin{itemize}\itemsep0em
\item in intra-postures validation, the trained CNN is evaluated on the fold not used for training;
\item in inter-posture validation, the model is evaluated on the data of a different posture, to measure the model's ability to generalize between postures.
\end{itemize}
All the results, divided by training posture pair, are shown in Figure \ref{img_bpPdrop} and in Table \ref{table_bpPdrop}.

With regard to intra-postures validation, the overall accuracy is $(\mu \pm \text{SE}) = (94.3 \pm 0.2)\%$, with $\sigma = 3.5\%$, which differs only by 0.2\% from the intra-posture validation accuracy obtained with single-session (thus single-posture) training (which is $(\mu \pm \text{SE}) = (94.5 \pm 0.2)\%$). This difference is comparable to the SEs of the two compared averages, thus not statistically significant. Moreover, the intra-postures validation accuracy yielded by two-posture training is higher than the corresponding baseline value of 90\% (standard deviation or SE not available). This indicates that in intra-posture validation the 1d-CNN performs better than the RBF-SVM.

The interpretation of these results is that the 1d-CNN, thanks to its higher capacity compared to SVM, is able to learn the hand gesture patterns coming from two arm postures with the same accuracy as it learns the patterns from a single posture. However, this success is marginal since the real aim is to improve inter-posture accuracy.

With regard to inter-posture validation, the overall inter-posture validation accuracy is 82.0\%, corresponding to a 12.3\% accuracy drop compared to the intra-postures case. The corresponding baseline value is 83\%, which is higher but does not allow for a statistical comparison since it is available without standard deviation or SE. This result is better by 1.4\% compared to the inter-posture validation accuracies achieved with single-session training. The accuracy drop quantifies the amount of overfitting produced by training on P$1+$P$i$, with respect to the task of generalizing to non-training postures. This amount of overfitting is slightly smaller compared to one given by the single-session (this single-posture) training, which was 13.9\%.

Thus, two-posture training improves the inter-posture generalization of the CNN. The fact that the intra-posture performance is not impacted, means that an amount of overfitting is removed without adding significant bias. This can be attributed to the 1d-CNN's capacity, which enables it to learn more diverse distributions (i.e.\ patterns from two postures instead that one) while preserving classification accuracy.

\begin{table}[H]
\centering
\small
\begin{tabular}{@{}ccccccc@{}}
\toprule
\multirow{2}{*}{\textbf{Train postures}} & \multicolumn{2}{c}{\textbf{Intra-postures accuracy}} & \multicolumn{2}{c}{\textbf{Inter-posture accuracy}} & \multicolumn{2}{c}{\textbf{Accuracy drop}} \\
 & $\mu \pm \text{SE}$ & $\sigma$ & $\mu \pm \text{SE}$ & $\sigma$ & $\mu \pm \text{SE}$ & $\sigma$ \\ \midrule
P$1+$P2 & $(94.2 \pm 0.2)\%$ & 3.2\% & $(82.4 \pm 0.4)\%$ & 7.3\% & $(11.8 \pm 0.4)\%$ & 6.4\% \\
P$1+$P3 & $(94.5 \pm 0.2)\%$ & 3.7\% & $(81.0 \pm 0.5)\%$ & 7.9\% & $(13.5 \pm 0.5)\%$ & 7.1\% \\
P$1+$P4 & $(94.2 \pm 0.3)\%$ & 3.8\% & $(82.6 \pm 0.5)\%$ & 7.0\% & $(11.6 \pm 0.4)\%$ & 6.7\% \\ \bottomrule
\end{tabular}
\caption{Two-posture training: intra-postures and inter-posture validation accuracies.}
\label{table_bpPdrop}
\end{table}

\section{Multi-day training strategy}

\begin{figure}[H]
  \centering
  \begin{subfigure}[b]{0.75\textwidth}
  \includegraphics[width=\textwidth]{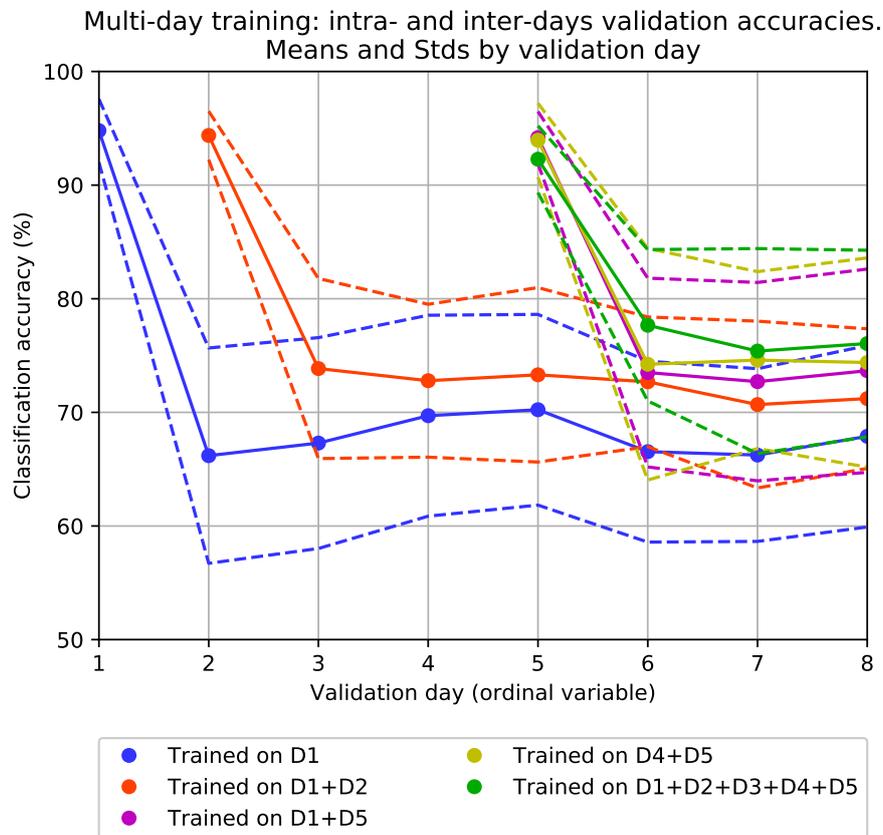}
  \end{subfigure}
  \caption{Multi-day training strategies compared to training on D1 alone: intra-day(s) accuracies and inter-day accuracies. Dots indicate average accuracies, and dashed lines indicate standard deviations.}
  \label{img_days_std}
\end{figure}

\begin{figure}[H]
  \centering
  \begin{subfigure}[b]{0.75\textwidth}
  \includegraphics[width=\textwidth]{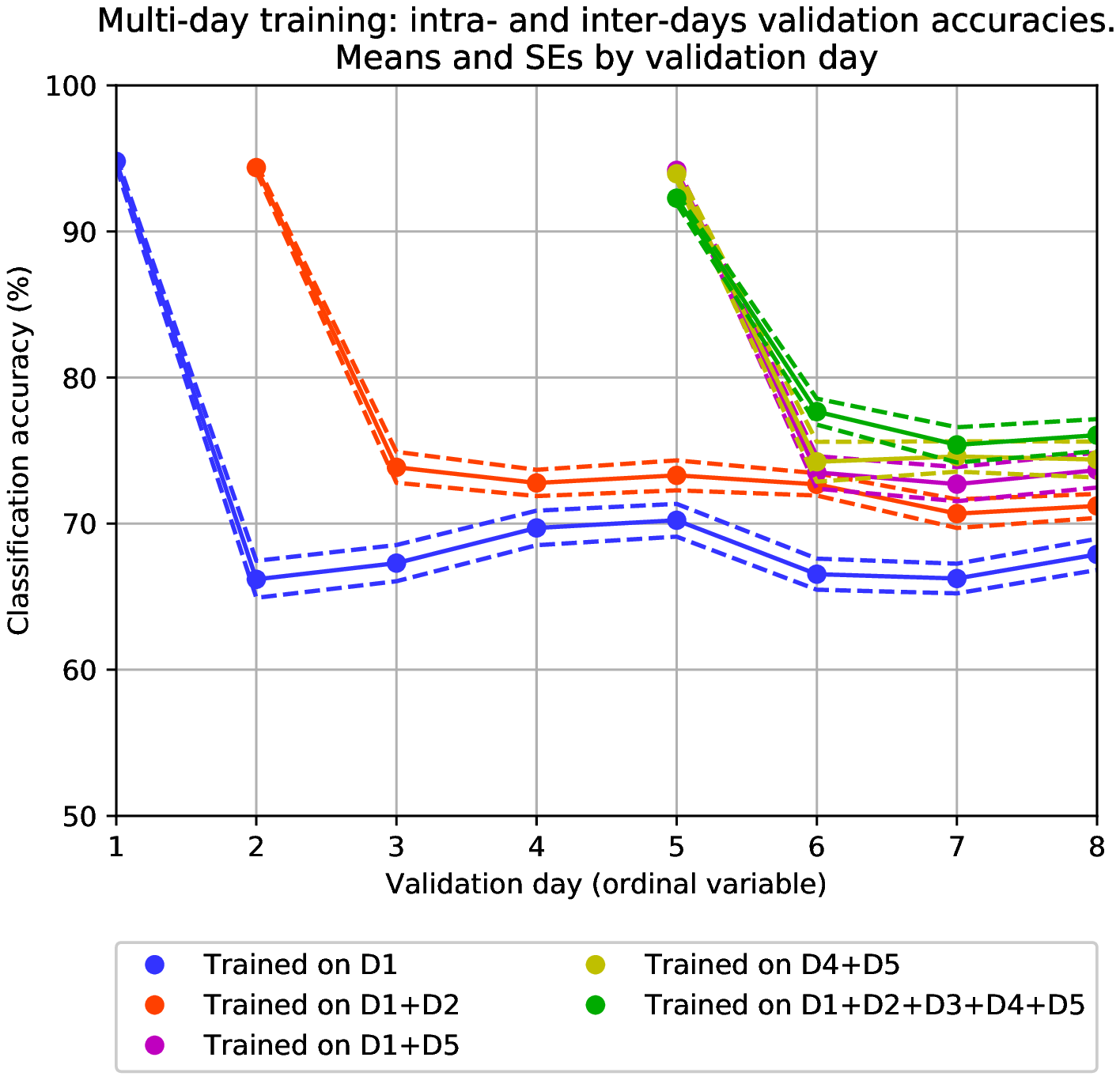}
  \end{subfigure}
  \caption{Multi-day training strategies compared to training on D1 alone: intra-day(s) accuracies and inter-day accuracies. Dots indicate average accuracies, and dashed lines indicate SEs.}
  \label{img_days_SE}
\end{figure}

The training strategy proposed to address temporal variability is multi-day training: for each subject, day combination, and arm posture, 2 CNNs are trained (one for each fold), using only data from that subject, those days and that posture. In particular, multi-day trainings are two-day trainings, which use D$1+$D2, D$1+$D5, and D$4+$D5, and five-day training, which use D1 to D5 (these combinations were chosen in order to repeat the setup which produced the baseline results). Performance is then assessed via intra-days validation and inter-day validation:
\begin{itemize}\itemsep0em
\item in intra-days validation, the trained CNN is evaluated on the fold not used for training;
\item in inter-day validation, the model is evaluated on the data of days D6 to D8 (absent in all multi-day training combinations), to measure the model's ability to generalize to different days.
\end{itemize}
All the results, divided by training posture pair, are shown in Figures \ref{img_days_std} and \ref{img_days_SE} and in Table \ref{table_days}, which also include a comparison with single-day (thus single-session) training on D1 alone.

With regard to inter-day validation of the training strategy based on D1 alone, the $(66.9 \pm 1.1)\%$ accuracy means a 27.9\% drop in accuracy compared to the intra-day validation, which yields $(94.8 \pm 0.3)\%$ accuracy. This proves that the inter-day variability is a larger effect than inter postural variability, whose impact was quantified as a 13.9\% accuracy drop (inter-posture validation of the single-session training strategy).

Improvement compared to single-day training is evident for all the multi-day training strategies implemented. The D1-to-D5 training strategy yields the best inter-day validation accuracy, $(\mu \pm \textrm{SE}) = (76.4 \pm 1.2)\%$ with $\sigma = 9.0\%$. The five-day training strategy proves to be the best one also in the baseline results based on classical machine learning algorithms. The corresponding baseline value is 77\%, which is consistent with the CNN result within the SE, allowing to say that the performance is statistically equivalent.

The second best strategy is D$4+$D5 strategy, achieving $(\mu \pm \textrm{SE}) = (74.4 \pm 1.2)\%$ with $\sigma = 9.2\%$. The statistical difference between the two was checked via a paired samples Wilcoxon test, which is the non-parametric equivalent of the paired samples $t$-test, chosen for its robustness with respect to sample distributions. The test yielded $p_\textrm{Wilcoxon} = 4.0\cdot 10^{-4}$, which means that the five-day training strategy is statistically significantly better (in validation) than the other ones.

Moreover, a trend can be noted which can be identified as user adaptation (explained in Subsection \ref{useradaptation}): the training strategies based on later days yield higher classification accuracy. This can be interpreted as the fact that inter-day differences in gesture execution decrease over time, as a consequence of the tendency of users to adapt to the repetitive exercise. This indicates again that training strategies prioritizing the recent data yield better performances.

\begin{table}[H]
\centering
\begin{tabular}{@{}ccccc@{}}
\toprule
\multirow{2}{*}{\textbf{Train days}} & \multicolumn{2}{c}{\textbf{Intra-days val. accuracy}} & \multicolumn{2}{c}{\textbf{Inter-day val. accuracy on D6 to D8}} \\
 & $\mu\pm\text{SE}$ & $\sigma$ & $\mu\pm\text{SE}$ & $\sigma$ \\ \midrule
D1 & $(94.8 \pm 0.3)\%$ & 2.3\% & $(66.9 \pm 1.1)\%$ & 8.0\% \\
D1+D2 & $(94.7 \pm 0.3)\%$ & 2.1\% & $(71.5 \pm 1.0)\%$ & 7.3\% \\
D1+D5 & $(94.2 \pm 0.3)\%$ & 2.3\% & $(73.3 \pm 1.2)\%$ & 8.9\% \\
D4+D5 & $(93.9 \pm 0.4)\%$ & 3.3\% & $(74.4 \pm 1.2)\%$ & 9.2\% \\
D1 to D5 & $(92.3 \pm 0.4)\%$ & 2.9\% & $(76.4 \pm 1.2)\%$ & 9.0\% \\ \bottomrule
\end{tabular}
  \caption{Multi-day training strategies compared to training on D1 alone: intra-day(s) accuracies and inter-day accuracies.}
  \label{table_days}
\end{table}

\section{Training strategies selection and test}

On the basis of the inter-session validation results, two-posture training was selected as the best strategy for postural generalization, and five-day training was selected as the best strategy for temporal generalization. This indicates that training strategies should prioritize data from more than one posture and from recent days. This outcome is the same as in the baseline results obtained with classical machine learning algorithms. After retraining, i.e.\ repeating the CNNs training using both folds, these two strategies were tested on the 10\% of data previously held out as test set (holdout details in Subsection \ref{datapartition}).

The two-posture training strategy yielded inter-posture test accuracy $(\mu \pm \text{SE}) = (81.2\pm 0.4)\%$ with $\sigma = 7.3\%$,. The five-day training strategy yielded inter-day (on D6 to D8) test accuracy $(\mu \pm \text{SE}) = (75.9\pm 0.7)\%$ with $\sigma = 8.6\%$

Test baseline values are not available, since the test step is not present in the pipeline which produced the baseline results. As reference, it is possible to consider the corresponding validation accuracies, which are 83\% for the inter-posture validation of the two-posture training and 77\% for the inter-day validation of the five-day training (both provided without $\sigma$ or SE). These results are similar to the test accuracies, but the limit of this comparison is that the baseline values, being the validation accuracies of the winner model (i.e.\ the RBF-SVM), are upward biased. However, the available values are sufficient to conclude that the final CNN performance is comparable with the baseline.

%Table \ref{table_test} sums up validation, test and baseline accuracies.

%\ begin{table}[]
%\ begin{tabular}{@{}cccccc@{}}
%\toprule
%\multirow{2}{*}{Train strategy} & \multicolumn{2}{c}{Inter-session val. accuracy} & \multicolumn{2}{c}{Inter-session test accuracy} & \multirow{2}{*}{Baseline} \\
% & \mu\pm\SE & \sigma & \mu\pm\SE & \sigma &  \\ \midrule
%Two-posture &  &  & (81.2\pm 0.4)\% & 7.3\% & 83\% \\
%Five-day & (76.4 \pm 1.2)\% & 9.0\% & 75.9\pm 0.7\% & 8.6\% & 77\% \\ \bottomrule
%\end{tabular}
%\end{table}

% % % % % % % % % % % % % % % % % % % % % % % % % % % % % % % % % % % % % % % % % % % % % % % % % % % % % % % % % % % % % % % % % % % % % % % % % % % % % % % % % % % % % % % % % % % % % % % % % % % % % % % % % % % % % % % % % % % % % % % % % % % % % % % % % % % % % % % % % % % % % % % % % % % % % % % % % % % % % % % % % % % % % % % % % % % % % % % % %

\chapter{Conclusions and future work}\label{conclusions}

This master thesis is the first application of deep learning on the Unibo-INAIL dataset, the first public sEMG dataset exploring the variability between subjects, sessions and arm postures, by collecting data over 8 sessions of each of 7 able-bodied subjects executing 6 hand gestures in 4 arm postures. With the open-source deep learning platform PyTorch, it was possible to implement and test a 1d-CNN architecture trained with most recent strategies based on training set composition.

The single-session training strategy achieves 94.5\% intra-session validation accuracy, but deteriorates to 80.6\% in inter-posture validation and to 66.9\% (for day 1) in inter-day validation. This proves that inter-day variability has a larger impact than inter-posture variability. A possible reason is that, on each day, the data of the 4 arm postures were collected without repositioning the sensors.

Multi-posture and multi-day training strategies yield higher inter-session validation accuracies. Two-posture training proves to be the best postural training strategy, indicating that the training strategies should prioritize data from more than one posture, and yields 81.2\% inter-posture test accuracy. Five-day training proves to be the best multi-day training strategy, and yields 75.9\% inter-day test accuracy. All the results are close to the baseline, provided by the accuracy of a RBF-SVM.

Moreover, the results of multi-day trainings allow to highlight user adaptation, the phenomenon which causes the inter-day differences in gesture execution to decrease over time, due to the tendency of users to adapt to the repetitive exercise during the first days. The detection of user adaptation indicates that the training strategies should also prioritize recent data.

Though not better than the baseline, the achieved classification accuracies rightfully place the 1d-CNN among the candidates for further research.

Future work will continue this research line investigating whether the fact that the deep 1d-CNN does not outperform the baseline is preprocessing-dependent or is due to an accuracy limit inherent to the Unibo-INAIL dataset. The question will be addressed using deep learning models relying on different data pre-processing, the first candidate being time-frequency domain analysed with 2d-CNNs.

% % % % % % % % % % % % % % % % % % % % % % % % % % % % % % % % % % % % % % % % % % % % % % % % % % % % % % % % % % % % % % % % % % % % % % % % % % % % % % % % % % % % % % % % % % % % % % % % % % % % % % % % % % % % % % % % % % % % % % % % % % % % % % % % % % % % % % % % % % % % % % % % % % % % % % % % % % % % % % % % % % % % % % % % % % % % % % % % %

\chapter*{Ringraziamenti}
\addcontentsline{toc}{chapter}{Ringraziamenti}
%\addcontentsline{toc}{chapter}{\protect\numberline{}Ringraziamenti}

Al termine di questo lavoro, desidero rivolgere dei ringraziamenti.

Ringrazio il Prof.\ Daniel Remondini per la proficua supervisione durante il lavoro di tesi, e per l’incoraggiamento che mi ha fornito. Ringrazio il Dott.\ Simone Benatti, il Dott.\ Francesco Conti e il Dott.\ Manuele Rusci per la guida costante nel corso del lavoro di ricerca, e per il prezioso supporto che mi hanno dato. Ringrazio queste persone anche per i commenti, sempre dettagliati e tempestivi, elargiti nel corso della redazione dell’elaborato di tesi. Ringrazio infine il Prof.\ Luca Benini per avermi permesso di svolgere il lavoro di tesi magistrale presso il laboratorio Energy-Efficient Embedded Systems (EEES, Unibo) da lui coordinato.

% % % % % % % % % % % % % % % % % % % % % % % % % % % % % % % % % % % % % % % % % % % % % % % % % % % % % % % % % % % % % % % % % % % % % % % % % % % % % % % % % % % % % % % % % % % % % % % % % % % % % % % % % % % % % % % % % % % % % % % % % % % % % % % % % % % % % % % % % % % % % % % % % % % % % % % % % % % % % % % % % % % % % % % % % % % % % % % % %

%\nocite{*}
%\printbibliography

%\chapter*{Bibliography}
%\addcontentsline{toc}{chapter}{Bibliography}
%%\addcontentsline{toc}{chapter}{\protect\numberline{}Ringraziamenti}

\end{document}